\documentclass[acmsmall,screen]{acmart}

\AtBeginDocument{%
  \providecommand\BibTeX{{%
    \normalfont B\kern-0.5em{\scshape i\kern-0.25em b}\kern-0.8em\TeX}}}

\usepackage{listings}
\usepackage{booktabs}   
\usepackage{subcaption} 
\usepackage{algorithm}
\usepackage{algpseudocode}
\usepackage{proof,latexsym,bm}
\usepackage{xspace}
\usepackage{macro}
\usepackage{wrapfig}
\usepackage{float}
\usepackage{tikz}
\usepackage{multirow}
\usepackage{graphicx}
\usepackage{pifont}
\usepackage{minibox}
\usepackage{url}

\setcopyright{rightsretained}
\acmPrice{}
\acmDOI{10.1145/3591288}
\acmYear{2023}
\copyrightyear{2023}
\acmSubmissionID{pldi23main-p470-p}
\acmJournal{PACMPL}
\acmVolume{7}
\acmNumber{PLDI}
\acmArticle{174}
\acmMonth{6}
\received{2022-11-10}
\received[accepted]{2023-03-31}

\begin{document}

\title[]{Inductive Program Synthesis via Iterative Forward-Backward Abstract Interpretation} 

\author{Yongho Yoon}
\email{yhyoon@ropas.snu.ac.kr}
\orcid{0009-0005-4962-0416}
\affiliation{
  \institution{Seoul National University}
  \department{Dept. of Computer Science \& Engineering}
  \country{Korea}
}
\author{Woosuk Lee}
\authornote{Corresponding author}
\email{woosuk@hanyang.ac.kr}
\orcid{0000-0002-1884-619X}
\affiliation{
  \institution{Hanyang University}
  \department{Dept. of Computer Science \& Engineering}
  \country{Korea}
}
\author{Kwangkeun Yi}
\email{kwang@ropas.snu.ac.kr}
\orcid{0009-0007-5027-2177}
\affiliation{
  \institution{Seoul National University}
  \department{Dept. of Computer Science \& Engineering}
  \country{Korea}
}


\begin{abstract}
A key challenge in example-based program synthesis is the gigantic
search space of programs.  To address this challenge, various work
proposed to use abstract interpretation to prune the search space.
However, most of existing approaches have focused only on forward
abstract interpretation, and thus cannot fully exploit the power of
abstract interpretation.  In this paper, we propose a novel approach
to inductive program synthesis via iterative forward-backward abstract
interpretation.  The forward abstract interpretation computes possible
outputs of a program given inputs, while the backward abstract
interpretation computes possible inputs of a program given outputs.
By iteratively performing the two abstract interpretations in an
alternating fashion, we can effectively determine if any completion of
each partial program as a candidate can satisfy the input-output
examples.  We apply our approach to a standard formulation,
syntax-guided synthesis (SyGuS), thereby supporting a wide range of
inductive synthesis tasks.  We have implemented our approach and
evaluated it on a set of benchmarks from the prior work.  The
experimental results show that our approach significantly outperforms
the state-of-the-art approaches thanks to the sophisticated abstract
interpretation techniques.
\end{abstract}




\maketitle

\section{Problem and Our Approach}
\label{sec:intro}

Inductive program synthesis aims to synthesize a program that
satisfies a given set of input-output examples.
The popular top-down search strategy is to enumerate \emph{partial
  programs} with missing parts  
and then complete them to a full program.

Though such a strategy is effective for synthesizing small programs, 
it hardly scales to large programs without being able to 
rapidly reject spurious candidates due to the exponential size of the search space.

Therefore, various techniques have been proposed to prune the search space
~\cite{scythe,morpheus,Polikarpova:2016,flashfill,duet}.
In particular, abstract interpretation~\cite{RiYi2020,Cousot2022}
has been widely used for pruning the search space 
of inductive program synthesis~\cite{simpl,scythe,morpheus,storyboard,blaze}. 
There are two kinds of abstract interpretation:
forward abstract interpretation that simulates program executions 
and backward abstract interpretation that simulates \emph{reverse} executions.

Most of the previous methods are solely based on forward abstract interpretation. 
They symbolically execute each partial program using their abstract semantics 
and compute a sound over-approximation of all possible outputs of programs
derivable from the partial program. 
If the over-approximated output does not subsume the desired output,
the program can be safely discarded.  

However, forward abstract interpretation alone is not sufficient 
because 
it just tells us 
the \emph{feasibility} of a partial program, but not about how to complete it. 
Backward abstract interpretation, on the other hand,
can be used to derive \emph{necessary preconditions} for the missing
parts of a partial program.


In this paper, we propose a new abstract interpretation-based pruning
method for inductive program synthesis that uses both forward and backward
reasoning in an iterative manner. 
For each partial program with missing expressions, 
a forward analysis computes (over-approximated) invariants 
over the program's final outputs and the results of intermediary operations 
from the given input examples. 
Based on the result of the forward analysis and the desired output
examples, a backward analysis computes \emph{necessary preconditions}
that must be satisfied by the missing expressions in order for the
program to produce the desired output.  These two analyses are
synergistically combined in a way that the result of one analysis
refines the result of the other, and are iterated until convergence.
If any of the necessary preconditions cannot be satisfied, the partial
program is discarded because it cannot produce the desired output.

\begin{figure}[t]
\centering
\includegraphics[scale=0.37]{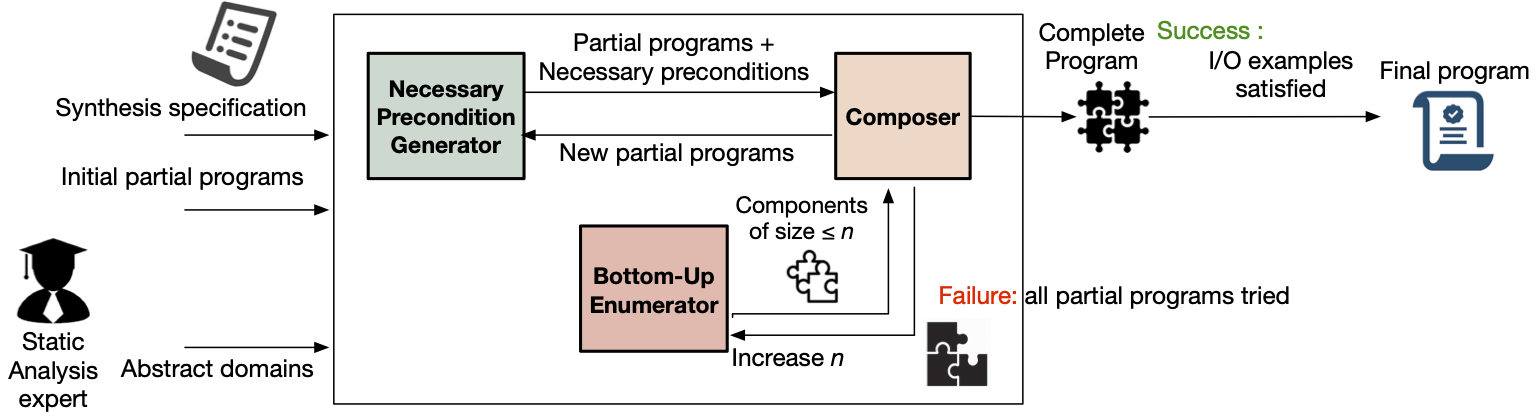}
\vspace{-0.1in}
\caption{High-level architecture of our synthesis algorithm.}
\vspace{-0.2in}
\label{fig:overview}
\end{figure}
    
Fig.~\ref{fig:overview} depicts the overall architecture of our synthesis algorithm, 
inspired by a recently proposed synthesis strategy~\cite{duet}. 
The algorithm takes synthesis specification comprising input-output 
examples, initial partial programs with missing parts as input. 
Additionally, it requires an abstract domain designed by domain experts 
that characterizes the abstract semantics of the target language.
Our algorithm consists of three key modules, namely 
\emph{Bottom-up enumerator}, \emph{Necessary precondition generator}, and 
\emph{Composer}:
\begin{itemize}
\item {\bf \emph{Bottom-up enumerator}}: Given input-output 
examples and a number $n$ which is initially $1$, 
the bottom-up enumerator generates \emph{components}. 
The components are expressions (of size $\leq n$) that are to be used to
complete the missing parts of partial programs. The bottom-up
enumerator exhaustively generates components of the size bound modulo
observational equivalence. 

\item {\bf \emph{Necessary precondition generator}}: Given a partial
  program with missing parts, 
the necessary precondition generator computes necessary preconditions
that must be satisfied by the missing expressions in order for the
program to satisfy the specification.  
To do so, it iteratively performs a forward and a backward abstract interpretations.
The resulting necessary precondition maps each missing expression to 
abstract values, which represent an over-approximation of the possible values
that the missing expression is allowed to generate in order for the 
program to satisfy the specification.
\item {\bf \emph{Composer}}: 
Given a partial program annotated with necessary preconditions 
and components,
the composer selects which hole to fill with which component
and generates a new partial program.
When putting a component in a missing part,
the composer checks if the necessary precondition of the missing part
is satisfied by the component.
If no component satisfies any of the necessary preconditions,
the partial program is discarded.
If a solution cannot be found until 
all the combinations of components and missing parts are tried,
the current set of components is determined to be insufficient. 
In this case, the bottom-up enumerator is invoked to add larger 
components (by increasing $n$), and the process is repeated.
\end{itemize}
Our algorithm is guaranteed to find a solution if it exists because the
\emph{Bottom-up enumerator} will eventually generate components to
complete the synthesis that satisfies the input-output examples. 

We have applied our approach to the SyGuS~\cite{sygus} specification language. 
SyGuS is a standard formation that has established various synthesis benchmarks 
through annual competitions~\cite{syguscomp}. 
SyGuS employs a formal grammar to describe the space of possible programs. 
Such a grammar is expressible in some SMT theory. 
We devise highly precise abstract domains 
specialized for the operators in theories of bitvectors and SAT. 
By targetting the standard formulation, our synthesis algorithm is applicable 
for a broad class of SyGuS problems with arbitrary grammars in those theories.

We implemented our algorithm in a tool called \tool. 
We evaluated \tool on a set of 5 benchmarks from the prior work 
on various applications: 
500 benchmarks from program deobfuscation \cite{David2020QSynthA},
369 benchmarks from program optimization \cite{lobster},
and 1006 benchmarks from the SyGuS competition (synthesizing
side-channel resistant circuits and bit-twiddling tricks) \cite{syguscomp}.
Our evaluation results show that \tool is more scalable than 
the state-of-the-art tools for inductive SyGuS problems 
\duet~\cite{duet} and \probe~\cite{probe}.
For example, for the 544 non-conditional bitvector-manipulation problems,
\tool is able to solve 519 problems in less than 34.8 seconds on average per problem, 
compared to only 456 and 409 by \duet and \probe using 165.3 and 63.7 seconds on average, respectively. 
\tool provides significant speedup over the state-of-the-art tools.

We summarize the main contributions of our work: 
\begin{itemize}
\item A novel and general synthesis algorithm that prunes the search
  space effectively by using both forward and backward abstract
  interpretation: Unlike existing synthesis algorithms, our algorithm
  uses both forward and backward reasoning thus fully exploit the
  power of abstract interpretation to prune the search space.
\item A highly precise abstract domain for bitvectors and SAT: We
  devise precise forward and backward abstract transfer functions for
  bitvector and boolean operators.  The resulting abstract domains are
  highly precise and can be used for inductive SyGuS problems with
  arbitrary grammars in the theories of bitvectors and SAT.
\item Implementation and evaluation of our algorithm: We implemented
  our algorithm in a tool called \tool and evaluated it on a set of 5
  benchmarks from a variety of applications. The results show
  significant performance gains over the existing state-of-the-art
  synthesis techniques.
\end{itemize}
    
\paragraph{Limitations.}
Our method requires a highly precise abstract domain for the target
application. 
In this paper, we have shown that our algorithm is effective for 
synthesizing bit-vector and Boolean expressions 
thanks to our highly precise abstract domains.  
However, whether our algorithm is effective for 
other applications (e.g., synthesizing programs with loops, 
or synthesizing string-manipulating programs)
remains an open question. 
We discuss possible directions for future work in Section~\ref{sec:futurework}.

\section{Overview}
\label{sec:overview}

We illustrate our method on the problem of synthesizing a bit-manipulating program. 
The desired program is a function $f$ that takes as input 
a bit-vector of fixed-width $4$ denoted $x$  
and turns off all bits left to the rightmost 0-bit in $x$.
Let us represent bit-vectors as binary numbers.
For example, given a bit-vector $1011_2$, 
the function is supposed to return $0011_2$. 

This problem is represented in SyGuS language, 
which formulates a synthesis problem as a combination of 
a syntactic specification and a semantic specification.
The syntactic specification for $f$ is the following grammar:
\[
\begin{array}{rclr}
S & \rightarrow & x \;|\; 0001_2 & \text{input bit-vector and bit-vector literals} \\
  &      \mid       & S \land S \;|\;  S \lor S \;|\; S \oplus S & \text{bitwise logical binary operators}\\
  &      \mid       & S+S \;|\; S \times S \; | \;S / S \; | \; S >\!> S & \text{bitwise arithmetic binary operators} 
\end{array}
\]
where $S$ is the start non-terminal symbol, and the operators are the ones supported 
in the theory of bit-vectors 
($\oplus$ denotes the bitwise exclusive-or operator and 
$>\!>$ denotes the arithmetic right shift operator where the leftmost bits are filled with the most significant bit of the left operand).
%
The semantic specification for $f$ follows the programming-by-example (PBE) paradigm 
and comprises input-output examples. 
For ease of exposition, we assume only one input-output example is given:
$ f(1011_2) = 0011_2$. 
%
A solution to the synthesis problem is 
$f(x) = ((x + 0001_2) \oplus x) >\!> 0001_2 $. 


\paragraph{Abstract Domain.}
In our abstract interpretations we use the following abstract domain
of bit-vectors to represent the abstract semantics of the program. 
The \emph{bitwise domain} 
is a set of elements each of which is a sequence of \emph{abstract bits} of length $4$. 
Each abstract bit has a value from the set $\set{0,1,\bot,\top}$ 
where $\top$ represents the unknown value and $\bot$ represents no value.
The domain has abstract operators for the bitwise logical and arithmetic operators. 
The abstract operators simulate the behaviors of the concrete operators 
in the abstract domain.
From now on, we denote each abstract operator in this domain by the corresponding concrete operator 
with a superscript $\#$.
For example, $1\top 1 0 \land^\# 00\top\top = 00\top 0$. 

\paragraph{Generation of Initial Partial Programs.}

We first generate a fixed set $Q$ of partial programs 
which have one or more non-terminal symbols as placeholders.
Starting from the start symbol $S$, 
we exhaustively generate all possible partial programs by 
applying the production rules of the grammar to non-terminal symbols 
up to a certain depth. 
In this example, for illustration purpose,  
we assume that the $Q$ set has three partial programs:
$$Q = \set{ (S \oplus x) >\!> 0001_2, (S/x) >\!> 0001_2, S_1 \times S_2 }$$


\paragraph{Component Generation.}


The component generator then generates a set $C$ of components by the bottom-up enumerative search, 
which maintains a set of complete programs and 
progressively generates new programs by composing existing ones. 
The set of components consists of expressions of size $\leq n$ 
where $n$ is the size upperbound which is initially $1$. 
This upperbound is increased by $1$ whenever the current set of components is insufficient to synthesize a solution. 
The number of components is potentially exponential to $n$, 
but we can reduce the number of components by exploiting observational equivalence of expressions.
For example, if $x$ is in the component set, $x \lor 0000_2$ is not added to the set because they are observationally equivalent.  
Because initially $n=1$, the component set is 
$$C = \set{x, 0001_2}.$$
These components are used to complete the missing parts (i.e., nonterminals) of the partial programs in $Q$ 
in the following composition phase. 


\paragraph{Derivation of Necessary Preconditions.}

Before we start the composition phase, we derive a necessary
precondition over each subexpression (including nonterminals) of the
partial programs in $Q$ using forward and backward analyses using the
abstract domain.  A necessary precondition is represented as an
element in the bitwise abstract domain.  The followings are the
derivation of necessary preconditions for the partial programs in $Q$.
\begin{itemize}
\item $S_1 \times S_2$ (necessary preconditions: $S_1 \mapsto
  \top\top\top\top, S_2 \mapsto \top\top\top\top$): We first perform a
  forward analysis on the partial program to obtain invariants over
  the program's final output and the results of intermediary
  operations.  The nonterminals can be replaced by any expressions, so
  the abstract output of both nonterminals is $\top\top\top\top$.
  Thus, the abstract output of the entire program is
  $\top\top\top\top$.  Now we check the feasiblity of the partial
  program by checking whether the abstract output of the partial
  program is consistent with the desired output example.  This check
  can be simply done by applying the meet operator ($\meet$) to the
  abstract output of the partial program and the abstraction of the
  desired output.  Because the result does not contain $\bot$
  ($\top\top\top\top \meet 0011 = 0011$), the partial program is
  feasible.  Now the backward analysis can be performed to obtain
  necessary preconditions over the non-terminals.  From the desired
  output $0011_2$, considering possible overflows in the
  multiplication, $S_1$ and $S_2$ can be any value.  Therefore, the
  necessary precondition of $S_1$ (and $S_2$) is $\top\top\top\top$.

\item $(S \oplus x) >\!> 0001_2$ (necessary precondition: $S \mapsto
  110\top$): By the forward analysis,
  the abstract output of $(S \oplus x)$ is $\top\top\top\top$, and the
  abstract output of $(S \oplus x) >\!> 0001_2$ is $\top\top\top\top
  >\!>^\# 0001 = \top\top\top\top$.  Now we check the feasiblity of the
  partial program.
  Becaues the result is not inconsistent with the desired output
  ($\top\top\top\top \meet 0011 = 0011$), the partial program is
  feasible.  Now the backward analysis is performed.  From the desired
  output $0011_2$, we can derive the necessary precondition over $(S
  \oplus x)$ as $011\top$ because $011\top >\!>^\# 0001 = 0011$.  The
  necessary precondition over $S$ is $110\top$ because, for the input
  $1011_2$ for $x$, $110\top \oplus^\# 1011 = 011\top$.

\item $(S / x) >\!> 0001_2$ (the program is infeasible): By the
  forward analysis, the abstract output of $(S / x)$ is
  $000\top$. That is because  
  from the fact that the
  maximum possible value of $S$ is $1111_2$ (which is $15$ in
  decimal) and the input $x$'s value is $1011_2$ (which is $11$ in decimal), 
  the possible values of $(S / x)$ are $0$ and $1$, which is represented as $000\top$ in the bitwise domain.
  The abstract output of $(S / x) >\!> 0001_2$ is $0000$ 
  because $000\top >\!>^\# 0001 = 0000$.  Now we check the feasiblity of
  the partial program.  Because the result is inconsistent with the
  desired output ($0000 \meet 0011 = 00\bot\bot$), the partial program is
  infeasible.
  
\end{itemize}
As shown in the above, 
the partial program $(S / x) >\!> 0001_2$ is determined to be infeasible.  
Only the other two partial programs will be considered in the composition phase.


\paragraph{Composition Process.}

Given the partial programs in $Q$ with necessary preconditions and the
set $C$ of components, 
the composer generates new (partial) programs by 
replacing non-terminal symbols in the partial programs with components.

The composer first chooses $(S \oplus x) >\!> 0001_2$.  It then
searches for a component $c$ in $C = \set{x, 0001_2}$ such that the necessary
precondition over $S$ is subsumed by the abstract output of $c$.
There is no such component because the necessary precondition
$110\top$ is not subsumed by neither of the abstract outputs of $x$
($1011$) and $0001_2$ ($0001$).  Therefore, the composer discards
the partial program.

The next partial program is $S_1 \times S_2$. 
Suppose the composer replaces $S_1$ with $x$, 
obtaining a new partial program $x \times S_2$. 
Whenever a new partial program is generated, 
the necessary precondition generator is invoked to derive necessary preconditions over the non-terminals.
Because the multiplication operation is modulo $2^4$,
computing the necessary precondition over $S_2$ amounts to 
solving the following equation: 
$$ 1011_2 \times y = 0011_2 \quad {\rm mod}\ 16 $$ 
where $y$ represents the unknown value of $S_2$.
Using the extended Euclidean algorithm, 
we can find the solution to the equation as $y = 1001_2$.
Thus, the necessary precondition over $S_2$ is $1001$.
Unfortunately, there is no component in $C$
whose abstract output subsumes the necessary precondition. 
Therefore, the composer discards the partial program.



Because the composer exhausts all the partial programs in $Q$ 
without finding a solution,
it invokes the component generator to generate more components.

Next, suppose the component generator generates components of size $\leq 3$ 
resulting in $C = \set{x, 0001_2, x+0001_2, \cdots}$. 

The composer revisits the partial program $(S_1 \oplus x) >\!> 0001_2$. 
Recall the necessary precondition over $S_1$ is $110\top$. 
Now the component $x+0001_2$ whose output is $1100_2$ 
satisfies the necessary precondition. 
Therefore, the composer replaces $S_1$ with $x+0001_2$
to obtain a new program $((x+0001_2) \oplus x) >\!> 0001_2$. 
After evaluation of the program, 
the composer finds that the program is correct and returns it as a solution.

The rest of the paper is organized as follows. 
Section \ref{sec:algo} introduces preliminary concepts and describes our overall synthesis algorithm. 
Section \ref{sec:abstract-domains} presents our abstract domains 
specialized for the theories of bitvectors and Boolean logic. 
Section \ref{sec:evaluation} presents our experimental results. 
Section \ref{sec:related} discusses related work.
Section \ref{sec:futurework} discusses future work.
Section \ref{sec:conclusion} concludes.

\section{Overall Synthesis Algorithm}
\label{sec:algo}


In this section, we formulate our method. 
We first introduce preliminary concepts including 
terms, regular tree grammars, and syntax-guided synthesis 
over a finite set of examples. 
We then present our generic synthesis algorithm, 
which is based on iterative forward-backward abstract interpretation.

\subsection{Preliminaries}


\paragraph{Term}
  A signature $\Sigma$ is a set of function symbols, where each $f
  \in \Sigma$ is associated with a non-negative interger $n$, the
  arity of $f$ (denoted $arity(f)$). 
For $n \geq 0$, we denote the set of all n-ary elements $\Sigma$ by $\Sigma^{(n)}$.
Function symbols of 0-arity are \emph{constants}. 
Let $V$ be a set of variables. 
The set $T_{\Sigma, V}$ of all $\Sigma$-terms over $V$ is inductively
defined; $V \subseteq T_{\Sigma, V}$ and 
$\forall n \geq 0, f \in \Sigma^{(n)}.~ t_1, \cdots, t_n \in T_{\Sigma, V}. ~ f(t_1, \cdots, t_n) \in T_{\Sigma, V}$.
A term can be viewed as a tree. 

\paragraph{Position}
The set of positions of term $s$ is a set $\pos(s)$ of strings over the alphabet of natural numbers, which is inductively defined as follows:
\begin{itemize}
\item If $s = x \in V$, $\pos(s) = \set{\epsilon}$.
\item If $s = f(s_1, \cdots, s_n)$, then $\pos(s) = \set{\epsilon} \cup \bigcup^n_{i=1}\set{ip \mid p \in \pos(s_i)}$.
\end{itemize}
The position $\epsilon$ is called the root position of term $s$. 
For $p \in \pos(s)$, the subterm of $s$ at position $p$, denoted by $s\mid_{p}$, is defined by 
(i) $s \mid_\epsilon = s$ and (ii) $f(s_1, \cdots, s_n) \mid_{iq} = s_i \mid_q$.
For $p \in \pos(s)$, we denote by $s[p \gets t]$ the term that is obtained from $s$ by replacing the subterm at position $p$ by $t$. Formally, 
$s[\epsilon  \gets t] = t $ and $f(s_1, \cdots, s_n)[iq  \gets t] = f(s_1, \cdots, s_i[q \gets s], \cdots, s_n)$.

\paragraph{Regular Tree Grammar.} 
A regular tree grammar is a tuple $G = (N, \Sigma, S, \delta)$ where
$N$ is a finite set of nonterminal symbols (of arity 0), $\Sigma$ is a
signature, $S \in N$ is an initial nonterminal, and $\delta$ is
a finite set of productions of the form $A_0 \to \sigma^{(i)}(A_1,
\cdots, A_i)$ where each $A_j \in N$ is a nonterminal.
Given a tree (or a term) $t \in \term$, applying a production $r = A
\to \beta$ into $t$ replaces an occurrence of $A$ in $t$ with the right-hand side $\beta$. 
A tree $t \in \term$ is generated by the grammar $G$ iff it can be obtained by applying 
a sequence of productions $r_1, \cdots, r_n$ to the tree of which root
node represents the initial nonterminal $S$. 
All the trees that can be derived from $S$ are called the language of $G$ and denoted by $L(G)$. 



\paragraph{Inductive Syntax-Guided Synthesis}
The syntax-guided synthesis problem~\cite{sygus} is a tuple $\tup{G, \specs}$. 
The goal is to find a program $P$ 
that satisfies a given specification $\specs$ in a decidable theory. 
Programs must be written in a language $L(G)$ described by a regular-tree grammar $G$. 
In particular, we say a SyGuS instance is inductive if 
the specification is a set of input-output pairs 
$\specs = \bigcup^m_{j=1}\set{i_j \mapsto o_j} $ where $i_j$ and $o_j$ are values
\footnote{
SyGuS instances with a logical formula, 
which have a unique output that satisfies the formula for a given input, 
can be transformed into ones with input-output examples by counterexample-guided inductive synthesis (CEGIS)~\cite{sketch}.
}. 
Assuming a deterministic semantics $\sem{P}$ is assigned to each program $P$ in $L(G)$,
the goal is to find a program $P$ such that $\sem{P}(i) = o$ for all $i \mapsto o \in \specs$ (denoted $P \models \specs$).

\subsection{Overall Algorithm}

\begin{algorithm}[t]
    \small
  \caption{The \tool Algorithm} \label{alg:bidir}
   \begin{algorithmic}[1]
\Require A SyGuS instance $\tup{G, \specs = \bigcup^m_{j=1}\set{i_j \mapsto o_j}}$ where each $i_j,o_j \in D$
\Require Abstract domain $\hat{D}$ such that $\powerset{D} \galois{\gamma}{\alpha} \hat{D}$
\Require An integer $d$ for the maximum height of the partial programs 
\Ensure A solution program $P \in L(G)$ that satisfies $\specs$
\State $Q := \sketchgen(G, d)$ \label{alg:bidir:sketchgen}
\State $n := 1$  \label{alg:bidir:compsizeinit}
\State $\cm := \emptyset$ \label{alg:bidir:compinit}
\Repeat \label{alg:bidir:loophead}
    \State $Q' := Q$ \label{alg:bidir:initq}
    \State $\cm := \compgen(G, \cm, n)$  \Comment{$\cm : N \to \powerset{L(G)} $}  \label{alg:bidir:compgen}
    \While {$Q'$ is not empty} \label{alg:bidir:while}
        \State remove $P$ from $Q'$ \label{alg:bidir:remove}
        \If {\textsf{IsComplete}($P$) and $P \models \specs$} \label{alg:bidir:complete}
            \Return $P$ \label{alg:bidir:solfound}
        \Else
            \State $\mathcal{A} := \analyze(P, \specs)$ \Comment{$\mathcal{A} : \pos(P) \to \hat{D}^m$} \label{alg:bidir:analyze}
            \State \textbf{if} $\exists p \in \pos(P).~ \bot_{\hat{D}} \in \mathcal{A}(p)$ \textbf{then} \textbf{continue} \label{alg:bidir:continue2}
            \State $pos := \pick(\textsf{Holes}(P))$  \Comment{$pos \in \pos(P)$} \label{alg:bidir:pick} 
            \For {$c \in \cm(P\mid_{pos})$ s.t. $\tup{\alpha(\sem{c}(i_1)), \cdots, \alpha(\sem{c}(i_m))} \po \mathcal{A}(pos)$} \label{alg:bidir:foreach}
                \State $Q' := Q' \cup \{P[pos \gets c ]\}$ \label{alg:bidir:enqueue}
            \EndFor \label{alg:bidir:endforeach} 
        \EndIf
    \EndWhile \label{alg:bidir:endwhile}
    \State $n := n + 1$ \label{alg:bidir:increasen}
\Until {false} \label{alg:bidir:loopend}
\end{algorithmic}
\end{algorithm}

Now we describe our algorithm of bidirectional search-based inductive synthesis 
accelerated by using forward/backward abstract interpretation. 
Algo.~\ref{alg:bidir} shows the high-level structure of our synthesis algorithm, 
which takes 
\begin{itemize}
\item Inductive SyGuS instance with a regular tree grammar $G$ and $m$ input-output examples $\specs$ 
\item Abstract domain $\hat{D}$ that abstracts the set $D$ of values of all terms in $L(G)$ 
and a Galois connection $\alpha : \powerset{D} \to \hat{D}$ and $\gamma : \hat{D} \to \powerset{D}$
\item Integer $d$ that specifies the maximum height of the \emph{sketches} (partial programs with nonterminals) to be explored
\end{itemize}
The sketches of height $\leq d$ are enumerated top-down according to the grammar $G$, and 
inserted into the priority queue $Q$ (line \ref{alg:bidir:sketchgen}).
Then, the size upperbound $n$ for components is initially set to be $1$ (line \ref{alg:bidir:compsizeinit}). 
The integer $n$ will be increased by $1$ at each iteration (line \ref{alg:bidir:increasen}) 
until a solution is found. 
The component pool $\cm$ (which is initially the empty set) 
includes all the components of size $\leq n$ that are generated in a bottom-up fashion. 
The main loop (lines \ref{alg:bidir:loophead}--\ref{alg:bidir:loopend}) is repeated 
until a solution is found. 
The priority queue $Q'$ which will be used in a current iteration is initialized 
by inserting the sketches in $Q$ into $Q'$ 
(line \ref{alg:bidir:initq}).
The $\compgen$ procedure incrementally builds expressions of size $\leq n$ 
by composing the previously generated expressions (line \ref {alg:bidir:compgen}). 
By exploiting the \emph{observational equivalence}, 
$\cm$ does not include multiple components which are semantically equivalent to each other 
with respect to the specification. 
In the while loop (lines \ref{alg:bidir:while}--\ref{alg:bidir:endwhile}),  
the algorithm explores the search space determined by the current component pool $\cm$ and the set of sketches. 
If a candidate $P$ dequeued from $Q'$ is a complete program 
and correct with respect to the specification,  
$P$ is returned as a solution (line \ref{alg:bidir:solfound}). 
Otherwise, another candidate in $Q'$ is explored. 
If a candidate $P$ is a partial program, 
we analyze $P$ to infer a \emph{necessary precondition} for each subexpression in $P$. 
The $\analyze$ procedure computes a map $\mathcal{A}$ 
that maps each subexpression in $P$ to a tuple of $m$ abstract values in $\hat{D}$ (line \ref{alg:bidir:analyze}). 
The $j$-th abstract value represents a necessary precondition to be satisfied any expression 
that is substituted for the nonterminal in order to satisfy the $j$-th input-output example $i_j \mapsto o_j$.
If $P$ has a position having $\bot_{\hat{D}}$ representing no expression can be put in that position, 
$P$ is determined to be infeasible and discarded (line \ref{alg:bidir:continue2}). 
Otherwise, a nonterminal in $P$ is chosen (line \ref{alg:bidir:pick}).  
For each component $c$ that can be substituted for the nonterminal and 
satisfies the necessary precondition (line \ref{alg:bidir:foreach}), 
we replace the nonterminal with $c$ and 
enqueue the resulting program into $Q'$ (line \ref{alg:bidir:enqueue}).
If no solution is found with the current component pool $\cm$, 
the size $n$ is increased by $1$ (line \ref{alg:bidir:increasen}) 
and the main loop is repeated.

Our algorithm has the following properties. 
First, our algorithm is sound and complete in the following sense.
\begin{theorem}
    Algo.~\ref{alg:bidir} is sound and complete in the sense that 
    if a solution to a given inductive SyGuS instance exists, 
    Algo.~\ref{alg:bidir} eventually finds the solution.
\end{theorem}
Second, it can determine the unrealizability~\cite{10.1145/3385412.3385979} of a given inductive SyGuS instance 
if every sketch in the queue $Q'$ has a position having $\bot_{\hat{D}}$ and  
discarded at line \ref{alg:bidir:continue2} in the main loop. 
That is, no sketch can be completed to a feasible program.
Lastly, it can be solely used for top-down synthesis 
rather than bidirectional synthesis, 
which makes it applicable to existing top-down synthesis algorithms.
To do so, instead of using a fixed set of sketches, 
at each iteration 
of the main loop (lines \ref{alg:bidir:loophead}--\ref{alg:bidir:loopend}), 
we can add larger sketches to the queue $Q'$ 
by increasing the maximum height $d$ of the sketches to be explored 
while keeping the size upperbound $n$ to be $1$ 
until a solution is found.

\subsection{The Iterative Forward-Backward Analysis}

We describe the $\analyze$ procedure in Algo.~\ref{alg:bidir} in detail.
It is known that by alternating forward and backward analyses, 
we can compute an overapproximation of the set of states that are both 
reachable from the program entry and able to reach a desired state at the program exit~\cite{jlp92}.  
In our setting where a program is a tree, 
execution of a program starts at the leaves of the program tree 
(constants or the input variable) and proceeds to the root. 
Therefore, for each node of the program tree, 
a forward analysis computes an over-approximation of the set of 
values that may be computed from a subtree rooted at the node in a bottom-up manner. 
A backward analysis computes an over-approximation of the set of
values that may be used to generate output desired by its parent in a top-down fashion. 

Given a candidate $P$ and $n$ input-output examples $\specs$,  
the $\analyze$ procedure performs the iterative forward-backward analysis 
for each input-output example and combines the results to obtain a map $\mathcal{A}$.
The analysis result  $\mathcal{A}$ 
maps each position in $P$ to a tuple of abstract values $\hat{d}_1, \cdots, \hat{d}_m$. 

The forward abstract semantics with respect to an input-output example $i \mapsto o$ is characterized by 
the least fixpoint of the following function $\forward_{\tup{i, o}} : (\pos(P) \to \hat{D}) \to (\pos(P) \to \hat{D})$: 
\[
\begin{array}{cc}
    \forward_{\tup{i, o}} = \lambda X. ~ \mathcal{I}^{i}_\forward \join F^{\#}(X) &
    \text{where} \quad \mathcal{I}^{i}_\forward = \set{p \mapsto \left\{ \begin{array}{ll} \alpha(i) & (P\mid_p \in V) \\ \top & (P\mid_p \in N) \\ \alpha(P\mid_{p}) & (P\mid_{p} \text{ is a constant}) \\ \bot & (\text{otherwise}) \end{array} \right. \suchthat p \in \pos(P)}.
\end{array}    
\]
The initial state $\mathcal{I}^{i}_\forward$ maps each nonterminal in $N$ to $\top$ since a nonterminal represents a hole that can be filled by any expression.   
Each variable is mapped to the abstraction of the input example (i.e., $\alpha(i)$), 
and each constant is mapped to the abstraction of the constant itself.
The forward abstract function $F^{\#}$ is defined as follows:
\[\small
    F^{\#} (X) = \set{ p \mapsto 
    \left\{
    \begin{array}{ll}
    \overrightarrow{f}^\#(X(p1), \cdots, X(pk)) & (P\mid_{p} = f(\cdots), arity(f) = k) \\ 
    \bot & (\text{otherwise}) 
    \end{array} \right . \suchthat p \in \pos(P)
    } 
\]
where $\overrightarrow{f}^\#$ denotes the forward abstract operator of an operator $f$. 
It takes abstract values of arguments and returns a resulting abstract value.

The backward abstract semantics with respect to an input-output example $i \mapsto o$ is 
characterized by the greatest fixpoint of the following function 
$\backward_{\tup{i, o}} : (\pos(P) \to \hat{D}) \to (\pos(P) \to \hat{D})$:
\[
\begin{array}{cc}
    \backward_{\tup{i, o}} = \lambda X. ~ \mathcal{I}^o_\backward \meet B^{\#}(X) &
    \text{where} \quad \mathcal{I}^o_\backward = \set{p \mapsto \left\{ \begin{array}{ll} \alpha(o) & (p = \epsilon) \\ \top & (\text{otherwise}) \end{array} \right. \suchthat p \in \pos(P)}. 
\end{array}
\]
The final state $\mathcal{I}^o_\backward$ maps the root position to the abstraction of the output example (i.e., $\alpha(o)$) 
and every other position to $\top$.
The backward abstract function $B^{\#}$ is defined as follows:
\[\small
    B^{\#} (X) = \set{ p \mapsto 
    \left\{
    \begin{array}{ll}
    \overleftarrow{f}^\#_i (X(p'), X(p'1), \cdots, X(p'k) ) & (p = p'i, P\mid_{p'} = f(\cdots), arity(f) = k) \\ 
    \top & (\text{otherwise}) 
    \end{array} \right . \suchthat p \in \pos(P)
    } 
\]
where $\overleftarrow{f}^\#_i$ denotes the backward abstract operator of an operator $f$. 
It takes an abstract value of the result 
and abstract values of arguments, and returns an abstract value of the $i$-th argument, which corresponds to 
the \emph{necessary precondition} for the $i$-th argument to satisfy the input-output example.

The intersection of the forward and backward abstract semantics is computed by 
obtaining the limit of the following decreasing chain defined for all $n \in \mathbb{N}$ 
until the chain converges~\cite{jlp92}: 
$\dot{X}^0_{\tup{i, o}} = \m{\Meta{lfp}}\ \forward_{\tup{i, o}}$, 
$\dot{X}^{2n+1}_{\tup{i, o}} = \m{\Meta{gfp}}\ \lambda X. ~ \dot{X}^{2n}_{\tup{i, o}} \meet \backward_{\tup{i, o}}(X)$, and 
$\dot{X}^{2n+2}_{\tup{i, o}} = \m{\Meta{lfp}}\ \lambda X. ~ \dot{X}^{2n+1}_{\tup{i, o}} \meet \forward_{\tup{i, o}}(X)$. 

The \analyze procedure is defined as follows: 
\[\small
    \analyze(P, \bigcup^m_{j=1} i_j \mapsto o_j ) \!=\! 
    \set{ p \mapsto \tup{\mathcal{A}_1(p), \cdots, \mathcal{A}_m(p)}  \suchthat p \in \pos(P), \forall 1 \leq j \leq m.~\mathcal{A}_j \!=\! \lim_{n \to \infty} \dot{X}^n_{ \tup{i_j , o_j}}}
\]

\begin{example}
    Consider the partial program $P = (S \oplus x) >\!> 0001_2$ in the overview example in Section~\ref{sec:overview}  
    where the input example $i = 1011_2$ and the output example $o = 0011_2$. 
    The subterms of $P$ are $(S \oplus x)$, $0001_2$, $S$, and $x$, 
    and their positions are $1$, $2$, $11$, and $12$, respectively.
    Assuming that the abstract domain $\hat{D}$ is the bitwise domain, 
    the initial state for the forward analysis $\mathcal{I}^{i}_\forward$ is 
    $\set{ \epsilon \mapsto \bot\bot\bot\bot, 1 \mapsto \bot\bot\bot\bot, 2 \mapsto 0001, 11 \mapsto \top\top\top\top, 12 \mapsto 1011} $. 
    The initial state for the backward analysis $\mathcal{I}^{o}_\backward$ 
    maps the root position $\epsilon$ to $0011$ and every other position to $\top\top\top\top$.
    The table below shows the first three elements of the decreasing chain $\dot{X}^n_{\tup{i, o}}$: 
    \begin{center}
    \small
    \begin{tabular}{|c|c|c|c|}
    \hline
    Position $p$ & $\dot{X}^0_{\tup{i, o}}(p)$ & $\dot{X}^{1}_{\tup{i, o}}(p)$ & $\dot{X}^{2}_{\tup{i, o}}(p)$ \\ \hline
    $\epsilon$ & $\top\top\top\top$ & $0011$ & $0011$ \\ \hline
    $1$ & $\top\top\top\top$ & $011\top$ & $011\top$ \\ \hline
    $2$ & $0001$ & $0001$ & $0001$ \\ \hline
    $11$ & $\top\top\top\top$ & $110\top$ & $110\top$ \\ \hline
    $12$ & $1011$ & $1011$ & $1011$ \\ \hline
    \end{tabular}
    \end{center}
    We assume the forward and backward abstract operators in the bitwise domain 
    defined for the bitwise XOR operator and the arithmetic right shift operator 
    (described in Section~\ref{sec:abstract-domains}) are used. 
    Because the chain converges at $\dot{X}^2_{\tup{i, o}}$, 
    the $\analyze$ procedure returns it as the final result.

\end{example}

\subsection{Optimizations}
\label{ssec:optimization}

In the implementation, we apply the following optimizations
to improve the efficiency of Algo.~\ref{alg:bidir}.

\paragraph{Concretization to Avoid Unnecessary Scans.}
In addition to the component pool $\cm : N \to \powerset{L(G)}$ 
that memorizes the component expressions derivable from each nonterminal,
we also maintain 
an additional map ${\bf V} : N \to (D \times D) \to \powerset{L(G)}$ 
that returns components that exhibit 
a certain input-output behavior. 
For example, in the overview example in Section~\ref{sec:overview}, 
${\bf V}(S)(1011_2, 1100_2) = \set{x + 0001_2}$ 
because $x + 0001_2$ outputs $1100_2$ when given $1011_2$ as input.
This map is used to save the time for line \ref{alg:bidir:foreach} 
only when the analysis result $\mathcal{A}(pos)$ is precise enough 
(i.e., its concretization result is a small set).   
In such a case, instead of computing the set 
$\set{c \in \cm(P\mid_{pos} \suchthat \tup{\alpha(\sem{c}(i_1)), \cdots, \alpha(\sem{c}(i_m))} \po \mathcal{A}(pos))}$, 
we compute 
$\bigcap \set{{\bf V}(P\mid_{pos})(i_j, v_j) \suchthat \mathcal{A}(pos) = \tup{\hat{d}_1, \cdots, \hat{d}_m}, \forall 1 \leq j \leq m. ~ v_j \in \gamma(d_j) }$ 
to obtain the set of components that are compatible with the analysis result. 
This can save the time for line \ref{alg:bidir:foreach} 
by avoiding scanning the entire component pool $\cm$.

\paragraph{Divide-and-Conquer for Conditional Programs.}
In the case of synthesis of conditional programs,
we incorporate the divide-and-conquer enumerative approach~\cite{eusolver} 
into our algorithm as follows. 
First, for each input-output example, by using Algo.~\ref{alg:bidir}, 
we synthesize a conditional-free program which satisfies that example. 
Second, we combine these conditional-free programs into a single conditional program 
that works for all examples by using the previous decision tree learning algorithm from \cite{eusolver}. 
Conditional predicates are generated by bottom-up enumeration.

\section{Abstract Domains}
\label{sec:abstract-domains}

In this section, we propose efficient but precise abstract domains that can be used 
for a wide range of inductivey SyGuS problems with background theories of 
(1) fixed-width bitvectors and (2) propositional logic. 

For a space reason, 
we only present some noteworthy points of our abstract domains. 
The full details of our abstract domains are available in the supplementary material. 

\subsection{Notations}
For the theory of bitvectors of fixed-width $w$, 
we follow the standard syntax and semantics of the bit-vector operators 
described in the SMT-LIB v2.0 standard~\cite{smtlib2}. 
Unsigned and signed bit-vectors are represented as bitstrings of length $w$ (i.e., $\set{0,1}^w$).
The values of unsigned bit-vectors range over $\set{0, \cdots, 2^w - 1}$,
and those of signed ones range over $\set{-2^{w-1}, \cdots, 2^{w-1} - 1}$ respectively. 
We refer to $-2^{w-1}$, $2^{w-1} - 1$, $0$, and $2^w - 1$ as 
$\signedmin$, $\signedmax$, $\unsignedmin$, and $\unsignedmax$ respectively.
We denote $[n]_i$ as $i$-th bit of the bitstring representation of $n$ 
and $[n]_{i:j}$ as a substring of $n$ 
comprising $i$-th bit, $i$+1-th bit, $\cdots$, $j$-th bit of $n$.
The concatenation of two strings $n_1$ and $n_2$ is denoted by $n_1 \cdot n_2$, and 
the length of a string $n$ is denoted by $|n|$.
We denote $n[a/b]$ as the bitstring obtained by replacing every occurrence of $b$ in $n$ with $a$.
For an interval $[l,h]$, we denote $lb([l,h])$ and $ub([l,h])$ as the lower and upper bounds of $[l,h]$ respectively.
Lastly, for a bitstring $n$, we denote $\trailzeros(n)$ 
as the number of trailing zeros in $n$.

\subsection{Abstract Domain for Fixed-Width Bitvectors}

\paragraph{Abstract Domain.}
Our abstract domain for fixed-width bitvectors is a reduced product domain. 
Reduced product of abstract domains 
can be used to achieve precise abstractions by synergistically combining
the expressiveness power of several abstract domains~\cite{Cousot2022}.
Our abstract domain comprises the following domains. 

The \emph{bitwise domain} $\tup{\bitwise, \po_{\bitwise}, \join_{\bitwise}, \meet_{\bitwise}}$ is a domain that tracks the value of each bit of a bit-vector independently, 
also used in prior work~\cite{mine:hal-00748094,10.1145/1134650.1134657}.  
Each element of the bitwise domain $\bitwise$ is a string of abstract bits of length $w$ 
as already introduced in Section~\ref{sec:overview}. 
The domain is formally defined as follows:
    $\bitwise \defeq \set{ b_1 \cdot b_2 \cdots b_w \suchthat \forall 1 \leq i \leq w.~ b_i \in \set{0, 1, \bot, \top}} $.
To define the galois connection between $\bitwise$ and $\powerset{\Z}$, 
we define the following function that computes the bit-vector representation $p(x)$ of an integer $x$ using 
the two's complement representation~\cite{mine:hal-00748094}: 
\[\small
    p(x) \defeq \left\{ \begin{array}{ll}
        b_1 \cdots b_w \quad \text{ where } \forall 1\leq i \leq w. ~ b_i = \lfloor x / 2^{w-i} \rfloor \mod 2 & (x \geq 0)\\
        b'_1 \cdots b'_w  \quad \text{ where }  \forall 1\leq i \leq w. ~b'_i = \neg b_i , b_i = [p(-x-1)]_i &  (x < 0)\\
    \end{array} \right.
\]
Note that the leftmost bit is the most significant bit. 
The galois connection $\powerset{\Z}\galois{\gamma_{\bitwise}}{\alpha_{\bitwise}}\bitwise$ is 
defined as follows:
\[\small
\begin{array}{ccc}
    \alpha_{\bitwise}(Z) \defeq \bigsqcup_{\bitwise} \set{p(x) \suchthat x \in Z} & \quad &
    \gamma_{\bitwise}(b) \defeq \left\{ 
        \begin{array}{ll}
            \emptyset & (\exists i. ~ [b]_i = \bot) \\ 
            \gamma^{\rm unsigned}_{\bitwise}(b) \cup \gamma^{\rm signed}_{\bitwise}(b) & (\text{otherwise})
        \end{array}
    \right. 
\end{array}
\]
where the concretization functions for unsigned and signed bit-vectors are defined as follows:
\[\small
\begin{array}{c}
    \gamma^{\rm unsigned}_{\bitwise}(b) \defeq \set{ \sum^{w-1}_{i = 0} 2^i m_{w-i} \suchthat \forall 1 \leq i \leq w.~ m_i \in \gamma_B([b]_i) } \\    
    \gamma^{\rm signed}_{\bitwise}(b) \defeq \set{ -2^{w-1} m_1 + \sum^{w-2}_{i = 0} 2^i m_{w-i} \suchthat \forall 1 \leq i \leq w.~ m_i \in \gamma_B([b]_i) }.
\end{array}
\]

The \emph{signed interval domain} $\tup{\signed, \po_{\signed}, \join_{\signed}, \meet_{\signed}}$ is a domain for representing bitvectors as intervals of signed bit-vectors.
Formally,
    $\signed \defeq \set{ [l,h] \mid l, h \in \set{0,1}^w, \sem{\bvsle}(l,h) = {\tt true} }$
where $\bvsle$ is the binary predicate for signed less than or equal. 
The galois connection $\powerset{\Z}\galois{\gamma_{\signed}}{\alpha_{\signed}}\signed$ is 
standard. 

The \emph{unsigned interval domain} $\tup{\unsigned, \po_{\unsigned}, \join_{\unsigned}, \meet_{\unsigned}}$ is a domain for representing bitvectors as intervals of unsigned bit-vectors.
Formally,
    $\unsigned \defeq \set{ [l,h] \mid l, h \in \set{0,1}^w, \sem{\bvule}(l,h) = {\tt true} }$
where $\bvule$ is the binary predicate for unsigned less than or equal. 
The galois connection $\powerset{\Z}\galois{\gamma_{\unsigned}}{\alpha_{\unsigned}}\unsigned$ is also standard.

The above three domains are combined to form the product abstract domain $\hat{D} = \bitwise \times \signed \times \unsigned$.
with the galois connection $\powerset{\Z}\galois{\gamma}{\alpha}\hat{D}$ 
that can be simply defined by combining the galois connections of the three domains.

To let the information flow among the three domains to mutually refine them, 
we need to use a \emph{reduction} operator that exploits the information tracked by one of the three domains 
to refine the information tracked by the others. 
The reduction requires to compute a fixpoint~\cite{10.1007/3-540-56287-7_95},  
and our \emph{iterated} reduction operator $\rho : \hat{D} \to \hat{D}$ 
is defined as follows:
\[\small
    \rho \defeq \m{\Meta{fix}}\ \lambda \tup{b,s,u}. ~ \langle
        b \meet \pi_{\signed \to \bitwise}(s) \meet \pi_{\unsigned \to \bitwise}(u), 
        s \meet \pi_{\bitwise \to \signed}(b) \meet \pi_{\unsigned \to \signed}(u), 
        u \meet \pi_{\bitwise \to \unsigned}(b) \meet \pi_{\signed \to \unsigned}(s)
    \rangle
\]
where $\pi_{\hat{D}_1 \to \hat{D}_2}$ is our \emph{projection operator} 
that propagates information from abstract domain $\hat{D}_1$ to another domain $\hat{D}_2$.
For example, the $\pi_{\bitwise \to \unsigned}$ operator takes a bitwise element 
and returns an unsigned interval whose lower bound (resp. upper bound) is a bit-vector obtained by 
replacing every $\top$ abstract bit with 0 (resp. 1). 
More details can be found at the supplementary material.

A reduction operator $\rho$ in abstract domain $\hat{D}$ should be \emph{sound} 
in the sense that it has to satisfy the following two properties: for all $d \in \hat{D}$,  
(1) $\rho(d) \po d$ (the result of its application is a more precise abstract element); 
(2) $\gamma(\rho(d)) = \gamma(d)$ (an abstract element and its reduction has the same meaning).
\begin{theorem}
    Our reduction operator $\rho$ is sound. 
\end{theorem}

\paragraph{Abstract Operators.}
Now we define forward and backward abstract operators. 
Because we deal with binary numbers with a fixed number of digits, 
all the domains consider possible overflows/underflows. 
For each bit-vector operation $f$ of arity $k$, 
the forward abstract operator $\fsem{f} : \hat{D}^k \to \hat{D}$,  
which takes $k$ abstract elements of arguments and returns an abstract element of the result,  
is defined to be 
\[\small
    \fsem{f}(\tup{b_1, s_1, u_1}, \cdots, \tup{b_k, s_k, u_k}) = \rho (\tup{\fsem{f}_{\bitwise}(b_1, \cdots, b_k), \fsem{f}_{\signed}(s_1, \cdots, s_k), \fsem{f}_{\unsigned}(u_1, \cdots, u_k)})
\]
where $\fsem{f}_{\bitwise}$, $\fsem{f}_{\signed}$, and $\fsem{f}_{\unsigned}$ are the forward operators in the bitwise domain, the signed domain, and the unsigned domain, respectively.
For each $1 \leq i \leq k$, 
the backward abstract operator $\bsem{f}_i : \hat{D}^{k+1} \to \hat{D}^k$  
takes the current abstract element of the result and $k$ abstract elements of arguments,  
and returns a more refined abstract element of the $i$-th argument.
\[\small
    \bsem{f}_i(\tup{b, s, u}, \tup{b_1, s_1, u_1} \cdots \tup{b_k, s_k, u_k}) = \rho (\tup{\bsem{f}_{\bitwise, i}(b, b_1 \cdots b_k), \bsem{f}_{\signed, i}(s, s_1 \cdots s_k), \bsem{f}_{\unsigned, i}(u, u_1 \cdots u_k)})
\]
where $\bsem{f}_{\bitwise, i}$, $\bsem{f}_{\signed, i}$, and $\bsem{f}_{\unsigned, i}$ are the backward operators in the bitwise domain, the signed domain, and the unsigned domain, respectively.
We always apply the reduction operator whenever we apply the forward or backward operators 
in order to maintain analysis results in the most precise form. 

In the following, we describe the forward and backward operators for each domain. 
We will use the same bit-vector operator names as the ones defined in the SMT-LIB v2.0 standard.

\paragraph{Forward Abstract Operators for the Signed/Unsigned Interval Domains}

In the following, we explain some salient features of the forward abstract operators in detail.
\begin{itemize}
\item $\fsem{\bvadd}_{\signed}$ 
is defined similarly to the standard integer interval arithmetic 
with a slight difference that it is aware of possible overflows/underflows. 
Formally, 
\[\small
\begin{array}{rcl}
    \fsem{\bvadd}_{{\signed}}([l_1, h_1], [l_2, h_2]) &=& \wrap_{{\signed}}([\sem{\bvadd}(l_1, l_2), \sem{\bvadd}(h_1, h_2)]) \quad (\fsem{\bvadd}_{{\unsigned}}\text{ is similar.})    
\end{array}
\]
where $\wrap_{\signed}([l, h]) = [l, h]$ if ($[l, h] \po_{\signed} [\signedmin, \signedmax]$) 
and $[\signedmin, \signedmax]$ otherwise. 
The other abstract operators for subtraction, multiplication, and division are similarly defined.
\item 
$\fsem{\bvneg}_{\unsigned}$ switches the lower and upper bounds of the interval with 
the exception of the case where the lower bound represents zero and 
the upper bound represents a non-zero value. 
That is because the negation of $1$ is $-1$ 
which is represented as $11\cdots 1_2$ in two's complement representation.
$11\cdots 1_2$ is the largest value in the unsigned interval domain.
Therefore, the top element is returned in this case.
Formally, 
\[\small
\begin{array}{rcl}
    \fsem{\bvneg}_{\unsigned}([l, h]) &=& \left\{ 
        \begin{array}{ll}
        \top_{\unsigned} & (l = p(0), h \neq p(0) ) \\ [1ex]   
        \lbrack \sem{\bvneg}(h), \sem{\bvneg}(l) \rbrack & (\text{otherwise}) 
        \end{array} \right. \\ [1ex]   
\end{array}
\]
\item $\fsem{\bvsdiv}_{\signed}$ simulates the following concrete semantics of the signed division.
\[\small
    \sem{\bvsdiv}(s, t) = \left\{ \begin{array}{ll}
        \sem{\bvudiv}(s, t) & (\text{if $s$ and $t$ are both non-negative}) \\ 
        \sem{\bvudiv}(\sem{\bvneg}(s), \sem{\bvneg}(t)) & (\text{if both are negative}) \\ 
        \sem{\bvneg}(\sem{\bvudiv}(\sem{\bvneg}(s), t)) & (\text{if $s$ is negative and $t$ is non-negative}) \\
        \sem{\bvneg}(\sem{\bvudiv}(s, \sem{\bvneg}(t))) & (\text{if $s$ is non-negative and $t$ is negative})
    \end{array} \right. 
\]
Following the above semantics, $\fsem{\bvsdiv}_{\signed}$ splits the intervals of the arguments into the four cases
and computes the quotient of each case separately using the $\fsem{\bvudiv}$ operator, which is standardly defined.
\end{itemize}

\paragraph{Forward Abstract Operators for the Bitwise Domain}


Some remarkable points are as follows:
\begin{itemize}
\item $\fsem{\bvadd}_{\bitwise}$ is defined by the $\textsf{RippleCarryAdd}$ operator defined in \cite{10.1145/1134650.1134657},  
which simulates the ripple-carry addition of two bit-vectors in the bitwise domain. 
\item $\fsem{\bvneg}_{\bitwise}$ is defined from that $\sem{\bvneg}(b) = \sem{\bvadd}(\sem{\bvnot}(b), p(1))$~\cite{hackersdel}. 
\item $\fsem{\bvmul}_{\bitwise}$ is defined from that 
the number of trailing zeros in the result of $\sem{\bvmul}(b_1, b_2)$ is the sum of the number of trailing zeros in $b_1$ and $b_2$. 
Formally, 
\[
\begin{array}{rcl}
    \fsem{\bvmul}_{\bitwise}(b_1, b_2) &=&  \top^{w - (n+m)} \cdot 0^{(n+m)} 
    \quad \text{ where }n = \trailzeros(b_1), m = \trailzeros(b_2)  \\ [1ex]      
\end{array}
\]
\item The abstract operators for the arithmetic shift $\fsem{\bvashr}_{\bitwise}$ is defined 
using the shift-right operators in the bitwise domain defined as follows: 
$b \ashr_x i $ shifts all the abstract bits of $b$ to the right by $i$ bits,
and the leftmost bits are filled with $x$.
Formally, 
\[\small
\begin{array}{rcl}
    \fsem{\bvashr}_{\bitwise}(b_1, b_2) &=& \bigsqcup \set{b_1 \ashr_{[b_1]_1} (i \mod w) \suchthat i \in \gamma^{\rm unsigned}_{\bitwise}(b_2)}. 
\end{array}
\]
The other abstract shift operators are similarly defined. 
\end{itemize}

\paragraph{Backward Abstract Operators for the Signed/Unsigned Interval Domains}
Some noteworthy points are as follows:
\begin{itemize}
\item 
$\bsem{\bvurem}_{\unsigned,1}(i, i_1, i_2)$ 
refines the abstract value of the first operand ($i_1$)
from abstract values of the second operand ($i_2$) and the result ($i$), and is defined as follows: 
\[\small
\begin{array}{rcl}
    \bsem{\bvurem}_{\unsigned,1}(i, i_1, i_2) &=&  \left\{ 
        \begin{array}{ll}
            i \meet_{\unsigned} i_1 & (\sem{\bvule}(p(2^{w-1}), lb(i)) = {\tt true}) \\
            \top & (\text{otherwise})
        \end{array}
    \right. 
\end{array}
\]
This behavior is based on the fact that 
for some bit-vectors $b_1$, $b_2$, and $b$, 
if $\sem{\bvurem}(b_1, b_2) = b$ and the most significant bit of $b$ is 1, 
then $b = b_1$. 
The proof is available in the supplementary material. 
\end{itemize}

\paragraph{Backward Abstract Operators for the Bitwise Domain}
Some noteworthy operators are as follows:
\begin{itemize}
\item $\bsem{\bvand}_{\bitwise,1}$ infers the abstract value of the left operand of $\bvand$ from 
the abstract values of the other operand and the result. 
For example, for each abstract bit of the result, if the bit is $1$, 
we can infer that the corresponding bit of the first operand is $1$ as well. 
Formally, 
\[\small
\begin{array}{rcl}
    \bsem{\bvand}_{\bitwise,1}(b, b_1, b_2) &=& d_1d_2\cdots d_w 
    \quad \text{ where }  \forall 1 \leq i \leq w.~d_i = 
        \left\{ 
            \begin{array}{ll}
                1 & ([b]_i = 1)   \\
                0 & ([b]_i = 0, [b_2]_i = 1)   \\
                \top & (\text{otherwise})
            \end{array}
        \right.  
\end{array}
\]
The abstract operators for the other bitwise logical operators are defined similarly. 
\item 
For $\bsem{\bvshl}_{\bitwise,1}(b, b_1, b_2)$, 
if the shift amount is a constant, we can infer the abstract value of the first operand
by shifting $b$ to the right (i.e., reverse direction) by the shift amount. 
However, the shift amount is not exactly known in general, 
so we should overapproximate it. 
We apply the join operator into the results of shifting $b$ to the right by all possible shift amounts. 
The minimum shift amount is zero and
the maximum shift amount is the number of trailing zeros of $b$ obtainable 
assuming every unknown bit ($\top$) is $0$. 
This range can be more refined by considering the abstract value of the second operand. 
Formally, 
\[\small
\begin{array}{rcl}
    \bsem{\bvshl}_{\bitwise,1}(b, b_1, b_2) &=&  \bigsqcup \set{b \ashr_\top (n\ {\rm mod}\ w) \suchthat n \in \gamma_{\unsigned}([0, \trailzeros(b[0/\top])] \meet \pi_{\bitwise \to \unsigned}(b_2)) }
\end{array}
\]
\item 
$\bsem{\bvmul}_{\bitwise,1}$ is most complicated among the backward abstract operators, which 
is defined as follows:
\[\small
\begin{array}{l}
    \bsem{\bvmul}_{\bitwise,1}(b, b_1, b_2) =
    \left\{ 
        \begin{array}{ll}
            \textsf{InferMulOp}(b_2, b)  & (\size{\gamma^{\rm unsigned}_{\bitwise}(b_2)} = \size{\gamma^{\rm unsigned}_{\bitwise}(b)} = 1)\\
            \top^{w - l} \cdot 0^{l} & (\text{otherwise}) \\ 
        \end{array}
    \right.\\ [1ex]   
    \qquad \text{where } l = \max \set{0, \trailzeros(b) - \trailzeros(b_2[0/\top])}
\end{array}
\]
where the $\textsf{InferMulOp}$ will be defined soon. 
This backward abstract operator precisely infers the abstract value of the first operand 
if the second operand and the result are exactly known.
Suppose the second operand and the result represent non-negative numbers $n_2$ and $n$ respectively
\footnote{The bitwise multiplication is not aware of the signedness of the operands. 
Therefore, it is safe to consider the operands as non-negative numbers.}.  
Because the multiplication is modulo $2^w$, 
our goal is to find $x$ such that 
\begin{equation}
    \label{eq:mul}
    x \times n_2 \equiv n \quad (\textrm{mod} \ 2^{w})
\end{equation}
When both $n_2$ and $n$ are not zero,
we can find $x$ as follows (cases where $n_2$ or $n$ is zero are trivial):
let the numbers of trailing zeros of $b_2$ and $b$ are $t_2$ and $t$ respectively. 
Because $n_2 \neq 0 \land n \neq 0$, $n_2 = 2^{t_2} \times m_2$ and $n = 2^{t} \times m$ 
for some odd numbers $m_2$ and $m$.
We have two cases. 
\begin{itemize}
\item Case 1) $t \geq t_2$: 
We can transform the equation (\ref{eq:mul}) into 
$ x \times (n_2 / 2^{t_2}) \equiv (n / 2^{t_2}) \quad (\textrm{mod}\  2^{w - t_2}) $
because $n_2 / 2^{t_2}$ is odd, $n_2 / 2^{t_2}$ and $2^{w - t_2}$ are coprime. 
By the extended Euclidean algorithm, 
we can find the modular multiplicative inverse of $n_2 / 2^{t_2}$
modulo $2^{w - t_2}$. Let $y$ be the modular multiplicative inverse. Then, 
\[
\begin{array}{lr}
    x \times (n_2 / 2^{t_2}) \times y \equiv (n / 2^{t_2}) \times y \quad (\textrm{mod}\  2^{w - t_2}) & (\textrm{multiplying $y$ into both sides}) \\
    x \equiv (n / 2^{t_2}) \times y \quad (\textrm{mod}\  2^{w - t_2}) &  \\
    x \times 2^{t_2} \equiv (n / 2^{t_2}) \times y \times 2^{t_2} \quad (\textrm{mod}\  2^{w}) &  (\textrm{multiplying $2^{t_2}$ into both sides})\\
\end{array}
\]
In conclusion, in binary representation of $x$, 
the last $w - t_2$ bits must be equal to those of $n / 2^{t_2} \times y$. 
The first $t_2$ bits of $x$ can be any values.
\item Case 2) $t < t_2$: 
In this case, there is no solution to the equation (\ref{eq:mul}). 
That is because the linear congruence (\ref{eq:mul}) has solutions if and only if 
the greatest common divisor (gcd) of $n_2$ and $2^w$ divides $n$. 
The gcd of $n_2$ and $2^w$ is $2^{t_2}$, and it cannot divide $n$ because $t_2 > t$. 
\end{itemize}
The above case analysis is implemented in the function $\textsf{InferMultOperand}$ 
defined as follows: 
\[\centering
\textsf{InferMulOp}(b_2, b) = \left \{ 
    \begin{array}{ll}
    \top & (n_2 = n = 0) \\
    0 & (n_2 \neq 0, n = 0) \\
    \top^{t_2} \cdot [p(n/2^{t_2} \times y)]_{w-t_2:w} & (n_2 \neq 0, n \neq 0, t \geq t_2) \\
    \bot  & (n_2 = 0 \land  n \neq 0 \lor n_2 \neq 0 \land n \neq 0 \land t < t_2) \\
    \end{array}
\right. 
\]
where $n_2 = \gamma^{\rm unsigned}_{\bitwise}(b_2)$, $n = \gamma^{\rm unsigned}_{\bitwise}(b)$, 
$t = \trailzeros(b)$, $t_2 = \trailzeros(b_2)$, and $y$ is the modular multiplicative inverse of $n_2 / 2^{t_2}$ modulo $2^{w - t_2}$.

If the second operand and the result are not exactly known, to overapproximate the abstract value of the first operand,
we just exploit the fact that the number of trailing zeros of the result is 
the sum of the number of trailing zeros of the first operand and the second operand.

\end{itemize}

\subsection{Abstract Domain for Boolean Algebra}

Our abstract domain for Boolean algebra is based on the abstract domain for bit vectors. 
The abstract domain is the set $B = \set{0, 1, \bot, \top}$ and can be understood as 
a variant of the bitwise domain where the length $w = 1$. 
The abstract operators for 
Boolean operators such as \texttt{and}, \texttt{or}, \texttt{xor}, \texttt{not} 
are defined in the same way as in the bitwise domain.


\section{Evaluation}
\label{sec:evaluation}

We have implemented our approach in a tool called
\tool\footnote{{\bf S}ynthesis from {\bf I}nductive specification e{\bf M}powered by {\bf B}idirectional {\bf A}bstract interpretation}
which consists of 10K lines of OCaml code and employs Z3~\cite{z3} as the constraint solving engine.
Our tool is publicly available for download\footnote{https://github.com/yhyoon/simba}.

This section evaluates \tool to answer the following questions:

\begin{itemize}
\item[\textbf{Q1:}] How does \tool perform on synthesis tasks from a variety of different application domains?
\item[\textbf{Q2:}] How does \tool compare with existing synthesis techniques?
\item[\textbf{Q3:}] How effective is the abstract interpretation-based pruning technique in \tool 
for reducing the search space compared to other alternatives (e.g., using an SMT solver, no pruning)?
\end{itemize}

All of our experiments were run on a Linux machine with an Intel Xeon 2.6GHz CPU and 256GB of RAM.

\subsection{Experimental Setup}

\paragraph{Benchmarks.}

We chose {\bf 1,875} synthesis tasks from three different application domains:
i) bit-vector manipulation without conditionals (\textsc{BitVec}), 
ii) bit-vector manipulation with conditionals (\textsc{BitVec-Cond}), 
and iii) circuit transformation (\textsc{Circuit}),

They are from the benchmarks used for evaluating the \duet tool, 
prior work on program deobfuscation~\cite{David2020QSynthA},
and the annual SyGuS competition~\cite{syguscomp}.

The \textsc{BitVec} domain comprise 544 tasks of the background theory of 
bit-vector arithmetic. All of the solutions to these problems are condtional-free programs.
\begin{itemize}
\item \textsc{HD}: 44 benchmarks from the SyGuS competition suite (General track). 
These problems originate from the book \emph{Hacker's Delight}~\cite{hackersdel}, 
which is commonly referred to as the bible of bit-twiddling hacks.
The semantic specification is a universally-quantified first-order formula 
that is functionally equivalent to the target program.
\footnote{
We have slightly modified the original benchmarks, which are for 32-bit integers, to be for 64-bit integers.
This is for a fair comparison with \probe since \probe can handle 64-bit integers only.
}
\item \textsc{Deobfusc}: 500 benchmarks from the evaluation benchmarks of a program deobfuscator \textsc{QSynth} (dataset ``VR-EA" in \cite{David2020QSynthA}). 
These problems aim at finding programs equivalent to randomly generated bit-manipulating programs 
from input-output examples, and have been used to evaluate 
the state-of-the-art deobfuscators~\cite{David2020QSynthA,xyntia,syntia}.  
Because the obfuscated programs are syntactically complicated, 
the best practice so far is a \emph{black-box approach} that samples input-output behaviors 
from the obfuscated programs and synthesize the target programs from the samples.
Following this approach, we have randomly generated 20 input-output examples for each obfuscated program. 
\end{itemize}

The \textsc{BitVec-Cond} domain comprise 750 tasks from the \duet evaluation benchmarks
\footnote{The benchmarks are 
slight modifications of 
the SyGuS competition suite (PBE-BitVector track). 
The syntactic restriction in each problem is replaced by a more general grammar. 
}.
These problems concern finding programs equivalent to randomly generated bit-manipulating programs 
from input-output examples ranging from 10 to 1,000.  
In contrast to the \textsc{Deobfusc} benchmarks, 
the solutions to these problems often contain conditionals. 



The \textsc{Circuit} domain from the evaluation benchmarks of \duet 
comprise 581 tasks of the background theory of SAT. 
\begin{itemize}
    \item \textsc{Lobster}: 369 problems from \cite{lobster}. These
      problems are motivated by optimizing homomorphic evaluation
      circuits.  Each problem is, given a circuit $C$, to synthesize a
      circuit $C'$ that computes the same function as $C$ but has a
      smaller multiplicative depth that is functionally equivalent to
      $C$.
    \item \textsc{Crypto}: 212 problems used in the SyGuS competition
      and motivated by side-channel attacks on cryptographic modules
      in embedded systems.  Each problems is, given a circuit $C$, to
      synthesize a constant-time circuit $C'$ (i.e. resilent to timing
      attacks) that computes the same function as $C$.
\end{itemize}

\paragraph{Baseline Solvers.}
We compare \tool against existing general-purpose synthesis tools 
with some form of domain specialization. 
Our algorithm is generally applicable to any domain, 
but it requires a suitable abstract domain for the target problems 
to further improve the efficiency.
Thus, we compare \tool with the following general-purpose tools that 
employ a kind of domain specialization:
\begin{itemize}
\item \duet is the state-of-the-art tool for inductive SyGuS problems 
that employs a bidirectional search strategy with 
a domain specialization technique called 
\emph{top-down propagation}. 
Top-down propagation is a divide-and-conquer strategy that recursively 
decomposes a given synthesis problem into
multiple subproblems.
It requires \emph{inverse semantics operators} 
that should be designed for each usable operator in the target language.
\item \probe performs a bottom-up search with 
guidance from a probabilistic model. 
Such a probabilistic model can be learned 
\emph{just in time} during the search process 
by learning from partial solutions encountered along the way. 
Thus, such a model can be viewed as a result of domain specialization 
for each problem instance.
\end{itemize}
We compare \tool with \duet for all the benchmarks 
and with \probe only for the \textsc{BitVec} domain 
because the \textsc{Circuit} and the \textsc{BitVec-Cond} domains are beyond the scope of \probe
\footnote{
The solutions of the \textsc{BitVec-Cond} instances are large conditional programs  
that require extensive case splitting 
which is not supported by \probe.}. 


\subsection{Effectiveness of \tool}\label{ssec:eval_basic}

\begin{figure}
    \centering
    \begin{subfigure}[b]{0.32\textwidth}
        \includegraphics[width=\textwidth]{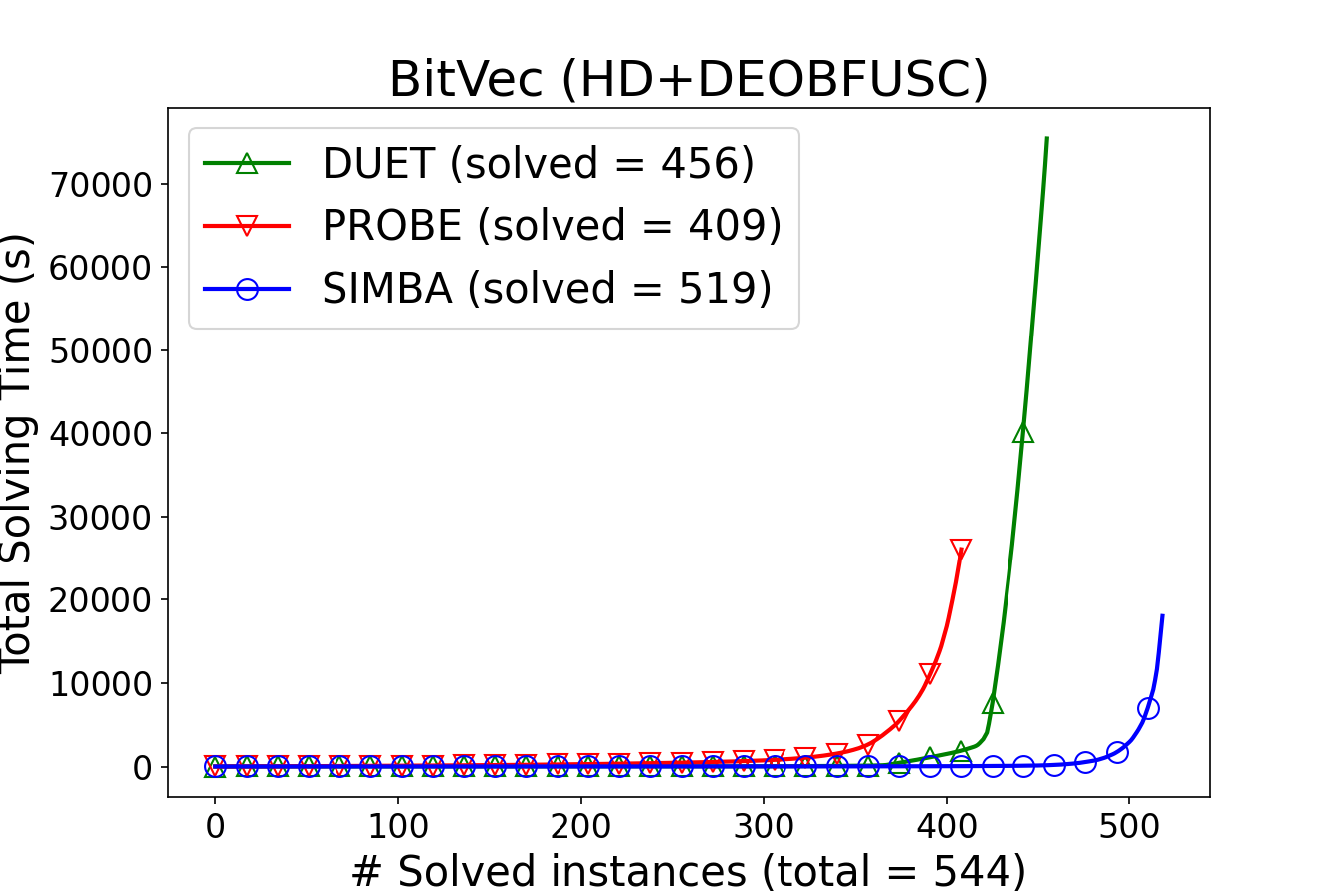}
    \end{subfigure}
    \hfill
    \begin{subfigure}[b]{0.32\textwidth}
        \includegraphics[width=\textwidth]{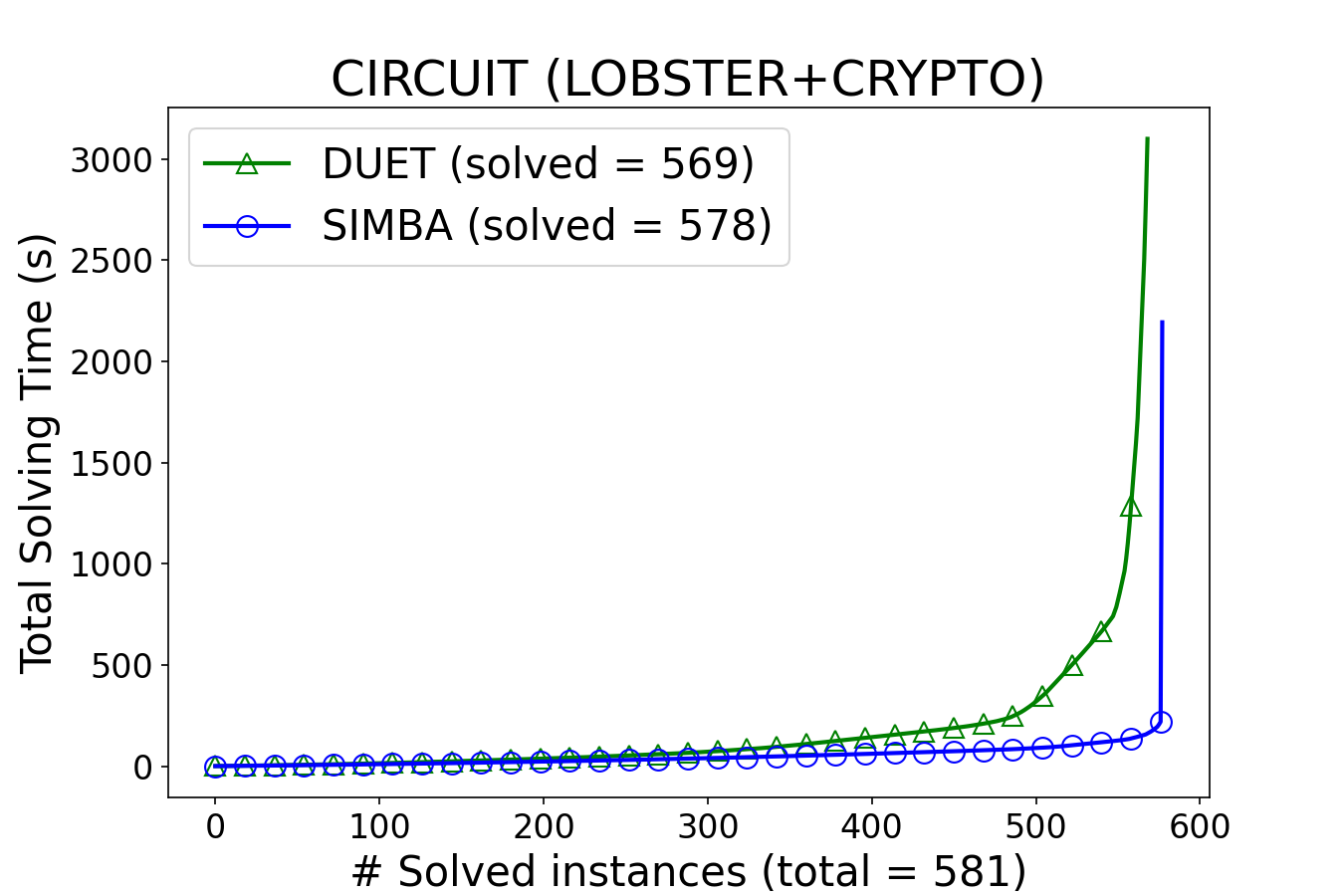}
    \end{subfigure}
    \hfill
    \begin{subfigure}[b]{0.32\textwidth}
        \includegraphics[width=\textwidth]{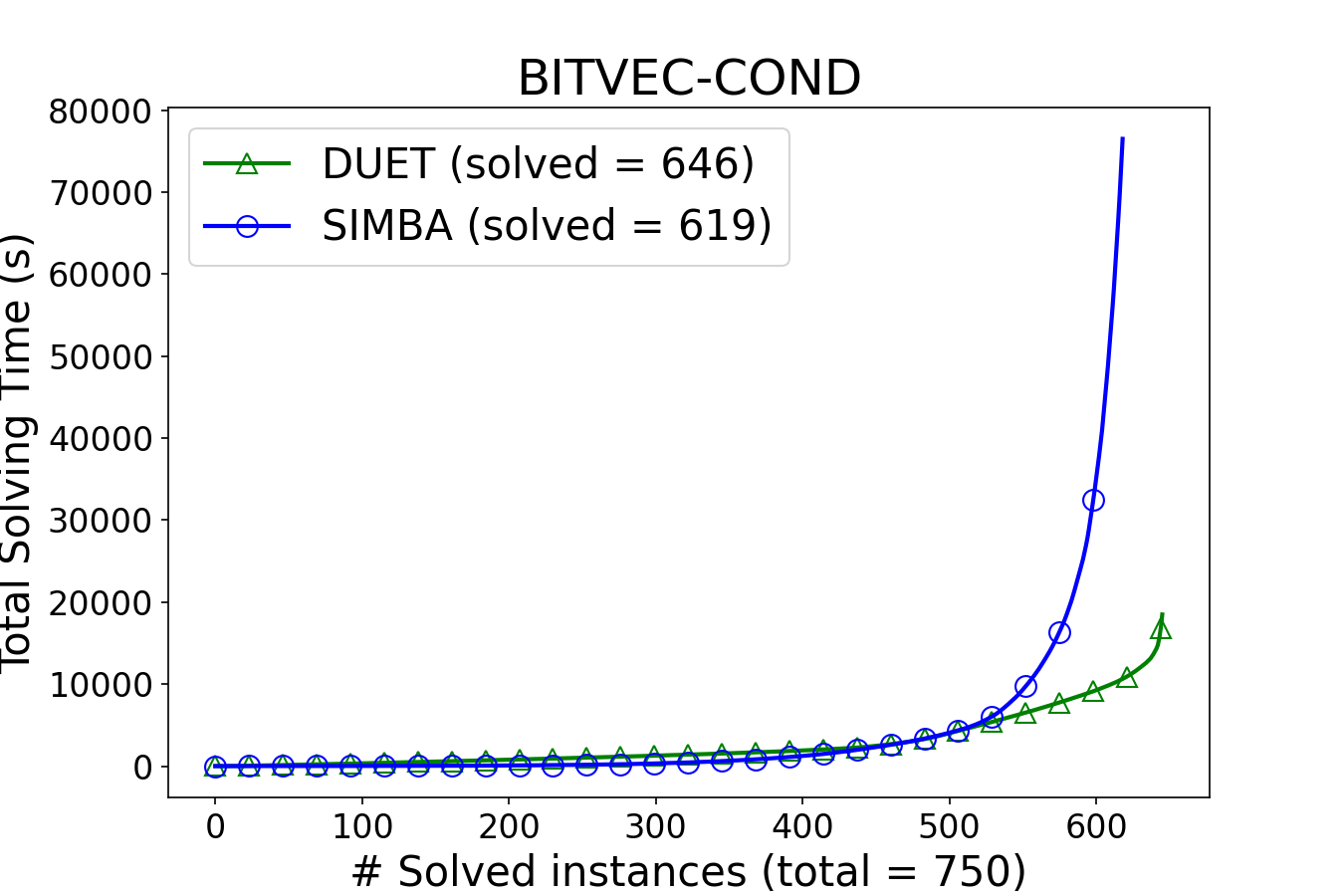}
    \end{subfigure}
    \caption{Comparison between \tool and the other baseline solvers on different domains.}
    \label{fig:cactus_tools}
\end{figure}

\begin{figure*}[t]
	\centering
    \begin{subfigure}[b]{0.32\textwidth}
        \includegraphics[width=\textwidth]{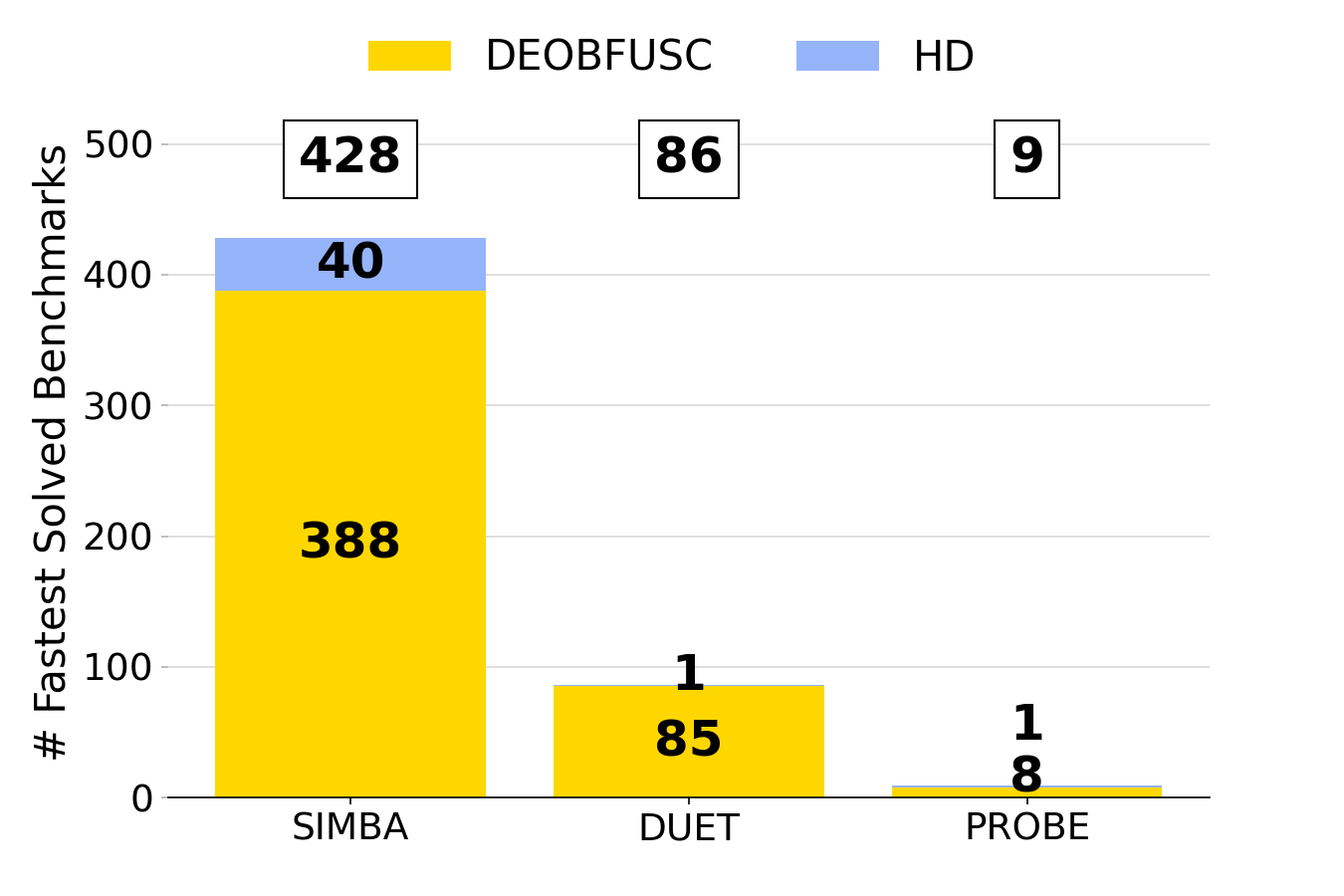}
        \caption{\# Fastest solved in the \textsc{BitVec} domain}
        \label{fig:bv_fast_count}
    \end{subfigure}
    \hfill
	  \begin{subfigure}[b]{0.32\textwidth}
        \includegraphics[width=\textwidth]{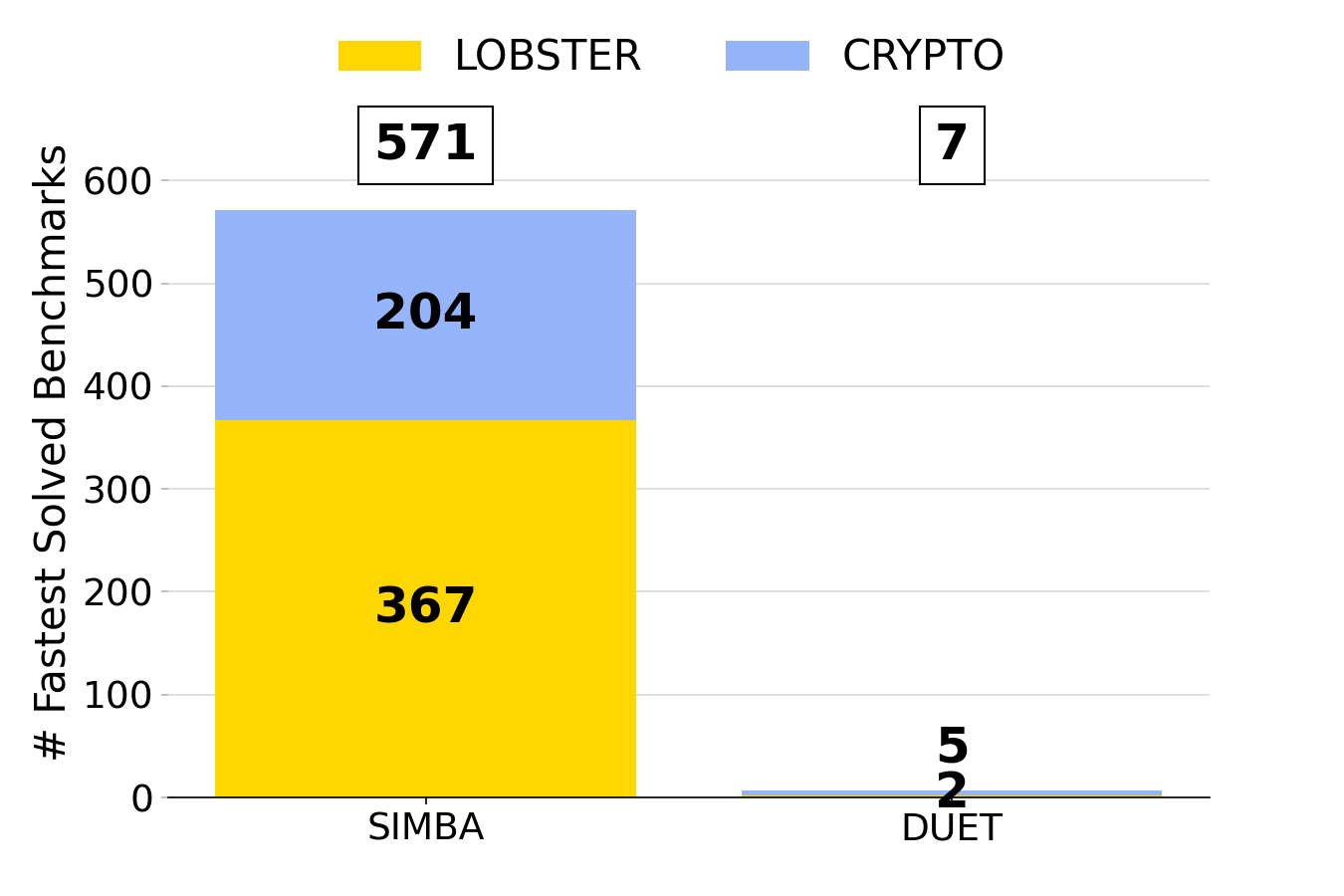}
        \caption{\# Fastest solved in the \textsc{Circuit} domain}
        \label{fig:circuit_fast_count}
    \end{subfigure}
    \hfill
    \begin{subfigure}[b]{0.32\textwidth}
      \includegraphics[width=\textwidth]{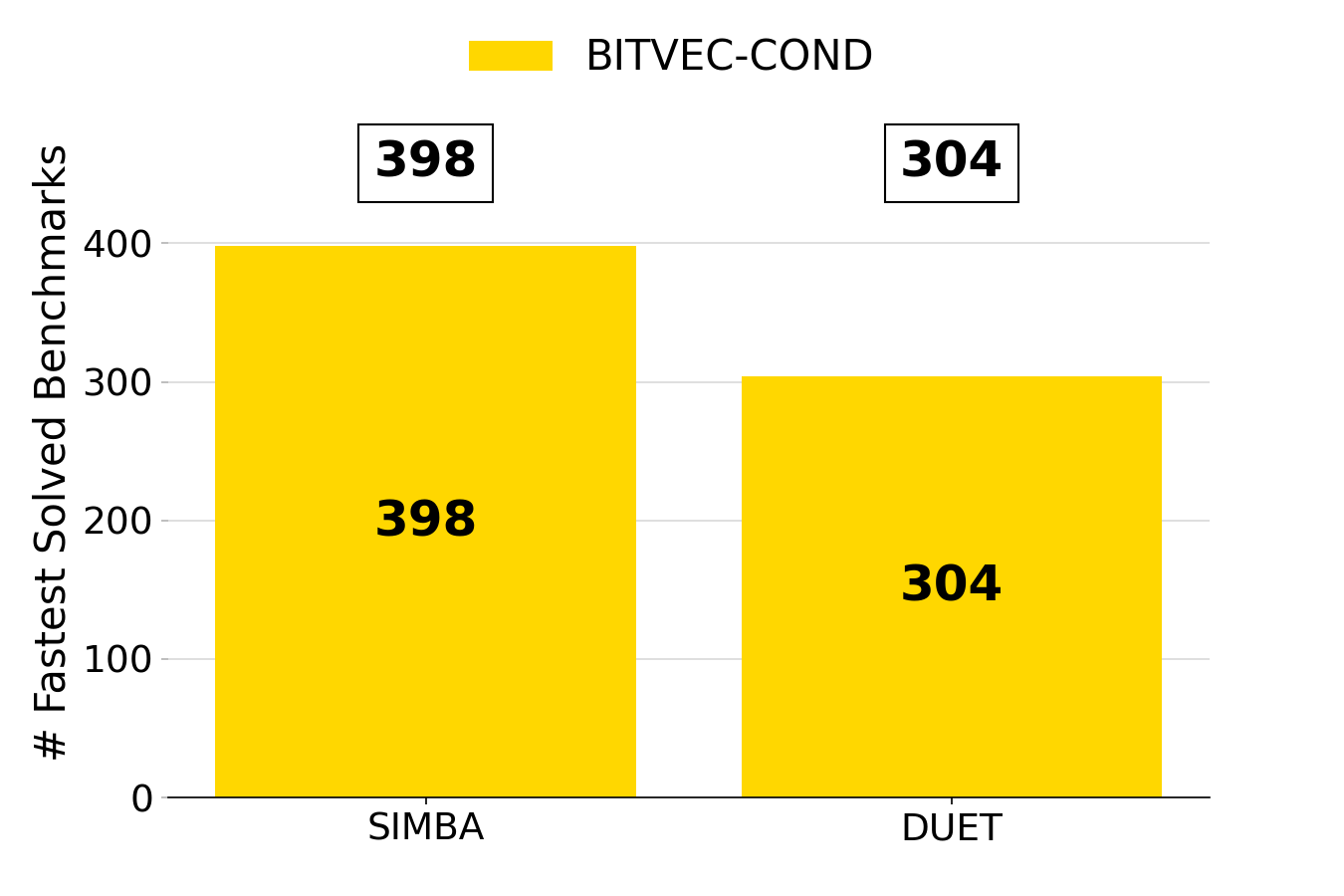}
      \caption{\# Fastest solved in the \textsc{BitVec-Cond} domain}
      \label{fig:bvcond_fast_count}
    \end{subfigure} \\
	  \begin{subfigure}{1.0\linewidth}
      \vspace{0.1in}    
      \resizebox{\textwidth}{!}{%
\small
\begin{tabular}{c|r|r|r|r|r|r|r|r|r|r|r|r|r|r|r} \hline
Benchmark  &
  \multicolumn{3}{c|}{\# Solved}	&
    \multicolumn{3}{c|}{Time (Average)} &
      \multicolumn{3}{c} {Time (Median)} &
        \multicolumn{3}{|c|}{Size (Average)} &
          \multicolumn{3}{c} {Size (Median)} \\ \cline{2-16}
category\! &
  \textsc{{\bf S}imba} & \textsc{{\bf D}uet} & \textsc{{\bf P}robe} &
    \textsc{{\bf S}} & \textsc{{\bf D}} & \textsc{{\bf P}} &
      \textsc{{\bf S}} & \textsc{{\bf D}} & \textsc{{\bf P}} &
        \textsc{{\bf S}} & \textsc{{\bf D}} & \textsc{{\bf P}} &
          \textsc{{\bf S}} & \textsc{{\bf D}} & \textsc{{\bf P}} \\
\hline \hline
\textsc{HD} \! &
  42 & 36 & 31 &
    5.7 & 42.7 & 11.8 &
      0.1 & 0.2 & 1.0 &
        7.7 & 8.5 & 6.5 &
          6 & 8 & 5 \\[0.3mm]
\textsc{Deobfusc} \! &
  477 & 420 & 378 &
    37.3 & 175.8 & 68.0 &
      0.0 & 0.1 & 2.9 &
        9.6 & 10.0 & 7.9 &
          9 & 9 & 8 \\[0.3mm]
\textsc{BitVec-Cond} \! &
  619 & 646 & - &
    123.6 & 28.6 & - &
      5.3 & 5.7 & - &
        269.6 & 350.6 & - &
          137 & 46 & - \\[0.3mm]
\textsc{Lobster}\! &
  369 & 369 & - &
    0.4 & 7.4 & - &
      0.2 & 0.8 & - &
        12.9 & 10.9 & - &
          13 & 11 & - \\[0.3mm]
\textsc{Crypto}\! &
  209 & 200 & - &
    9.7 & 1.8 & - &
      0.1 & 0.2 & - &
        11.9 & 12.8 & - &
          11 & 12 & - \\[0.3mm]
\hline
  {\bf Overall} \! &
  {\bf 1716} & {\bf 1671} & {\bf 409} &
    {\bf 56.4} & {\bf 58.0} & {\bf 63.7} &
      {\bf 0.2} & {\bf 1.7} & {\bf 2.7} &
        {\bf 107.1} & {\bf 145.4} & {\bf 7.8} &
          {\bf 15} & {\bf 13} & {\bf 8} \\[0.3mm]
\hline
\end{tabular}}
	      \caption{Statistics for the solving times and solution sizes.
                 All times are in seconds.
                 \probe could not run for the \textsc{Circuit} and \textsc{BitVec-Cond} domain. 
                 The number of problems in the
                 \textsc{HD}, \textsc{Deobfusc}, \textsc{Lobster}, \textsc{Crypto}, and \textsc{BitVec-Cond} categories
                 are {\bf 44}, {\bf 500}, {\bf 369}, {\bf 212}, and {\bf 750} respectively ({\bf 1875} in total).}	  
 	      \label{subfig:maintbl}
	  \end{subfigure} 
	  \vspace{-0.1in}
    \caption{Main result comparing the performance of \tool, \duet, 
             and \probe (breakdown by categories). The timeout is set to one hour.}
    \label{fig:main}
  \vspace{-0.2in}
\end{figure*}

We evaluate \tool on all the benchmarks and compare it with \duet and \probe.
For each instance, we measure the running time of synthesis and the size of the synthesized program, using a timeout of one hour.

The results are summarized in Fig.~\ref{fig:main}. 
Fig.~\ref{subfig:maintbl} shows the statistics of the solving times and solution sizes, 
and Fig.~\ref{fig:bv_fast_count} and \ref{fig:circuit_fast_count} 
show the number of instances solved with the fastest time for each domain per solver.
Because \probe is not applicable to the \textsc{Circuit} and \textsc{BitVec-Cond} domains, 
the results for \probe on these domains are not shown. 

Overall, \tool outperforms the other baseline tools both 
in terms of the number of instances solved and the average solving time.
As shown in Fig.~\ref{subfig:maintbl}, \tool solves 1716 instances,
while \duet and \probe solve 1671 and 409 instances respectively.
\tool is the fastest solver in 1397 instances (75\% of the total), 
while \duet is the fastest in 397 instances. 
Fig.~\ref{fig:cactus_tools}
shows the cactus plot of the solving times for \tool, \duet, and \probe.
The horizontal axis represents the number of solved instances
and the vertical axis represents the cumulative solving time.
The plot suggests that \tool solves more instances than \duet and \probe
in a shorter time. 

In comparison to \duet, \tool is more efficient in the \textsc{BitVec} domain 
and \textsc{Circuit} domain, 
while \duet is more efficient in the \textsc{BitVec-Cond} domain.
Since \tool performs the forward-backward analysis for each input-output example, 
the number of examples affects the efficiency. 
In the \textsc{BitVec-Cond} domain,
the number of examples is unusually large (up to 1000), which makes \tool inefficient.
On the other hand, \duet is able to handle the large number of examples efficiently. 
However, for the other domains (particularly \textsc{BitVec}), 
\tool outperforms \duet because \duet's inverse semantics 
often creates sub-problems that are hard to solve 
whereas \tool does not generate such sub-problems.




We measure solution quality by solution sizes in AST nodes. 
According to Occam's razor, smaller solutions are better since they are less likely to overfit the input-output examples. In general, \probe generates the smallest solutions (with average sizes of 6.5 and 7.9 for two domains), although \tool is also capable of generating solutions of similar sizes (with average sizes of 7.7 and 9.6). The slight gap between the two tools can be attributed to the fact that \tool can solve instances that \probe cannot solve due to its limited scalability. For example, \tool's solutions for the two domains not solved by \probe have average sizes of 10.5 and 13.1, respectively, while \tool's solutions to the ones solved by both tools have average sizes of 6.8 and 8.6, respectively.





\paragraph{Result in Detail.}

\begin{table}
  \footnotesize
  \centering
  \caption{Results for 25 randomly chosen benchmark problems (5 for each category), 
  where \textbf{Time} gives synthesis time,
  $T_{A}$ gives time spent for forward and backward analysis,
  and $|P|$ shows the size of the synthesized program (measured by number of AST nodes).}
  \vspace{-0.1in}
  \label{tbl:compare_detail}
  \begin{tabular}{c|l|rr|rr|rrr}
    \toprule
    Benchmark &
      \multirow{2}{*}{Benchmark} &
        \multicolumn{2}{c|}{ \probe } & \multicolumn{2}{c|}{ \duet } &
        \multicolumn{3}{c}{ \tool } \\
    category\! &  
      & 
        \textbf{Time} & $|P|$ & 
        \textbf{Time} & $|P|$ & 
        \textbf{Time} & $T_{A}$ & $|P|$  \\
    \hline
    \multirow{5}{*}{\textsc{HD}}
      & hd-03-d5-prog             &       0.85 &      4 &       1.19 &      4 &       0.06 & 0.00 &      4 \\
      & hd-07-d0-prog             &       0.92 &      6 &       0.09 &      6 &       0.07 & 0.00 &      6 \\
      & hd-14-d5-prog             &       4.91 &      9 &        >1h &      - &       0.60 & 0.26 &      9 \\
      & hd-19-d1-prog             &        >1h &      - &        >1h &      - &      10.79 & 7.07 &     19 \\
      & hd-20-d1-prog             &        >1h &      - &     228.98 &     15 &      98.22 & 0.22 &     15 \\
    \hline
    \multirow{5}{*}{\textsc{Deobfusc}}
      & target\_9                 &       >1h &      - &   2310.23 &     16 &     31.45 & 2.85 &     16 \\
      & target\_119               &      0.87 &      5 &      0.02 &      8 &      0.05 & 0.03 &      9 \\
      & target\_385               &     15.35 &      9 &     37.46 &     10 &      0.33 & 0.25 &      9 \\
      & target\_410               &       >1h &      - &   2635.78 &     16 &     69.38 & 3.06 &     16 \\
      & target\_449               &      2.61 &      7 &      0.01 &      7 &      0.03 & 0.00 &      7 \\
    \hline
    \multirow{5}{*}{\textsc{BitVec-Cond}}
      & 133\_1000                 &         - &      - &     48.18 &     35 &    166.55 & 6.89 &    139 \\
      & 23\_10                    &         - &      - &    340.75 &     62 &      0.89 & 0.40 &     15 \\
      & 60\_100                   &         - &      - &      6.90 &     14 &      0.15 & 0.10 &     14 \\
      & icfp\_gen\_10.7           &         - &      - &      4.68 &     67 &      1.19 & 0.20 &    192 \\
      & icfp\_gen\_14.1           &         - &      - &      6.29 &    154 &     49.61 & 37.75 &    415 \\
    \hline
    \multirow{5}{*}{\textsc{Lobster}}
      & hd09.eqn\_45\_0           &         - &      - &     21.95 &     15 &     11.09 & 0.32 &     13 \\
      & longest\_1bit-opt.eqn\_63\_1 &         - &      - &      0.59 &     13 &      0.17 & 0.01 &     11 \\
      & longest\_1bit-opt.eqn\_75\_1 &         - &      - &      1.31 &     14 &      0.24 & 0.01 &     15 \\
      & p03.eqn\_38\_2            &         - &      - &      0.22 &      9 &      0.11 & 0.01 &      9 \\
      & p09.eqn\_49\_1            &         - &      - &      0.24 &      7 &      0.19 & 0.01 &      9 \\
    \hline
    \multirow{5}{*}{\textsc{Crypto}}
      & CrCy\_2-P6\_2-P6          &         - &      - &    193.68 &     20 &      0.65 & 0.07 &     20 \\
      & CrCy\_5-P9-D5-sIn         &         - &      - &      0.13 &     11 &      0.11 & 0.00 &     11 \\
      & CrCy\_8-P12-D5-sIn4       &         - &      - &      0.21 &     11 &      0.14 & 0.01 &     11 \\
      & CrCy\_8-P12-D7-sIn5       &         - &      - &      1.99 &     19 &      0.63 & 0.04 &     19 \\
      & CrCy\_10-sbox2-D5-sIn11   &         - &      - &      0.28 &      9 &      0.13 & 0.00 &      9 \\
    \bottomrule
  \end{tabular}
  \vspace{-0.2in}
\end{table}

We study the results for each domain in detail. 
Table \ref{tbl:compare_detail} shows the detailed results on randomly chosen 25 problems 
(5 for each category).
The results suggest the significant impact of 
the forward and backward analyses. 
For example, for \texttt{hd-20-d1-prog}, 
\tool discarded $11,365$ out of total $11,710$ partial programs generated during the synthesis (97.1\%) 
with a small overhead of $0.22$ seconds.
For \texttt{target-410},
$187,858$ out of $188,501$ partial programs (99.7\%) are early pruned by \tool 
with an acceptable overhead of $3.06$ seconds.
Furthermore,
we observe that even partial programs are not discarded, 
the holes in the partial programs are highly constrained by 
the necessary preconditions, thereby 
significantly reducing the number of components to be explored 
for the completion of the partial programs.
On the other hand, 
\duet and \probe do not perform such pruning and 
thus generate many unlikely candidates, 
taking 10 to 1000 times longer than \tool. 

\paragraph{Analysis of the Impact of the Number of Examples.}
We study the impact of the number of examples on the performance of \tool.
Since the forward-backward analysis is performed as many times as the number of examples, 
the number of examples affects the efficiency, but 
the impact on efficiency is not significant in practice as long as 
the number of examples is not too large.  
We have conducted experiments with 500 deobfuscation benchmarks 
with the number of examples ranging from 5 to 20. 
When the number of examples given to \tool 
are 5, 10, and 20, 
the average synthesis time is 29.8, 31.0, and 37.3 seconds, respectively.
In all cases, the average solution size is 9.5. 
In other words, 
the average synthesis time increased by just 25\% 
when the number of examples increased fourfold (from 5 to 20).




\paragraph{Summary of Results.}
\tool solves harder synthesis problems more quickly compared to the
state-of-the-art baseline tools in diverse domains. 


\subsection{Efficacy of Our Abstract Interpretation-based Pruning}

\begin{figure*}[t]
	\centering
    \begin{subfigure}{0.32\linewidth}
        \centering
        \includegraphics[width=1.0\linewidth]{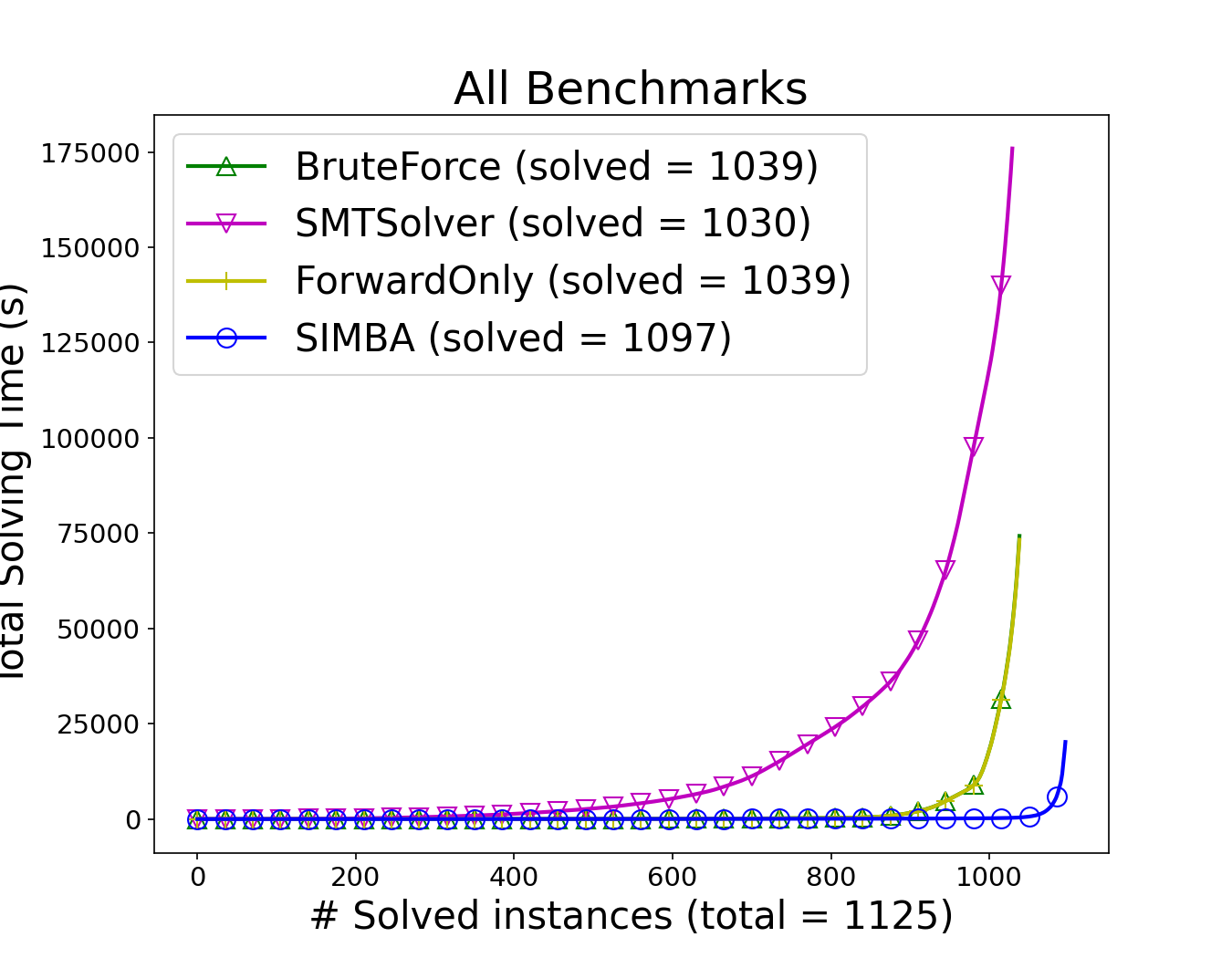}
	    \caption{Performance of different variants of \tool}
        \label{fig:cactus_ablation}
    \end{subfigure}
    \hfill
    \begin{subfigure}{0.64\linewidth}
        \vspace{0.1in}
        \centering
        \footnotesize
        \begin{tabular}{c|r|r|r|r|r|r|r|r}
\hline
Benchmark &
  \multicolumn{4}{c|}{\# Solved} &
    \multicolumn{4}{c}{Time (Average)} \\ 
        \cline{2-9}
category\! &
  {\bf S} & {\bf FW} & {\bf SMT} & {\bf BF} &
    {\bf S} & {\bf FW} & {\bf SMT} & {\bf BF} \\ 
\hline \hline
\textsc{HD} \! &
  42 &  41 & 40 & 41 &
    5.7 &  12.9 & 163.6 & 13.4 \\ [0.3mm] 
\textsc{Deobfusc}\! &
  477 &  421 & 424 & 421 &
    37.3 &  128.8 & 178.7 & 131.2 \\ [0.3mm] 
\textsc{Lobster}\! &
  369 &  368 & 358 & 368 &
    0.4 &  43.7 & 234.9 & 43.5 \\ [0.3mm] 
\textsc{Crypto}\! &
  209 &  209 & 208 & 209 &
    9.7 &  11.8 & 45.7 & 11.9 \\ [0.3mm] 
\hline
{\bf Overall}\! &
  {\bf 1097} &  {\bf 1039} & {\bf 1030} & {\bf 1039} &
    {\bf 18.4} &  {\bf 70.5} & {\bf 170.8} & {\bf 71.5} \\ [0.3mm] 
\hline
\end{tabular}
        \caption{Statistics for the solving times.
                 {\bf S}, {\bf FW}, {\bf SMT}, and {\bf BF} denote
                 \tool, \textit{ForwardOnly}, \textit{SMTSolver}, and \textit{BruteForce} 
                 variants respectively.}
        \label{subfig:ablation}
        \end{subfigure} 
	\vspace{-0.2in}
    \caption{Result for the ablation study.}
    \label{fig:ablation}
  \vspace{-0.2in}
\end{figure*}

We now evaluate the effectiveness of our abstract interpretation-based pruning.
For this purpose, we compare the performance of three variants of \tool, 
each using a different combination:
\begin{itemize}
\item \tool with the forward and backward analyses (i.e., the original \tool)
\item \textit{ForwardOnly} only with the forward analysis
\item \textit{BruteForce} without any pruning technique (i.e., only with the bidirectional search strategy)
\footnote{Despite the pruning is disabled, 
the other optimizations such as symmetry breaking and  
observational equivalence reduction are still enabled.}
\item \textit{SMTSolver} equipped with the SMT-based pruning. 
It checks the feasibility of each partial program 
by checking the satisfiability of an SMT formula. 
The formula encodes the partial program under construction 
and the desired input-output behavior of the program.
A similar approach was used in \morpheus~\cite{morpheus} to prune the search space. 
Because the SMT solving is without any approximation, 
it is expected to be more accurate than the abstract interpretation-based pruning 
at the cost of higher computational cost.
\end{itemize}
By comparing the performance of these variants, 
we aim to understand 
\begin{itemize}
\item the overall impact of our abstract interpretation-based pruning (\tool vs. \textit{BruteForce})
\item the necessity of the backward analysis (\tool vs. \textit{ForwardOnly} vs. \textit{BruteForce})
\item the cost-effectiveness of using an abstract interpreter (\textit{ForwardOnly} vs. \textit{SMTSolver})
\end{itemize}

Fig.~\ref{fig:ablation} shows the results of the ablation study 
(Fig.~\ref{subfig:ablation} shows the statistics for the solving times and
Fig.~\ref{fig:cactus_ablation} shows the cactus plots). 
We exclude the \textsc{BitVec-Cond} 
benchmarks which require extensive case-splitting 
dealt with by the divide-and-conquer strategy~\cite{eusolver} 
to focus on evaluating the core idea of our approach.

The first observation is that our abstract interpretation-based pruning
is effective in reducing the search space 
(1097 solved by \tool vs. 1039 solved by \textit{BruteForce}).
The second observation is that the backward analysis is necessary
to prune the search space effectively 
because \textit{BruteForce} and \textit{ForwardOnly} are almost the same 
whereas \tool is much better than \textit{ForwardOnly}.
The third observation is that using the SMT solver is more expensive than 
using an abstract interpreter 
(1039 solved by \textit{ForwardOnly} vs. 1030 solved by \textit{SMTSolver}).
Though the SMT solver is more accurate than the abstract interpreter, 
the success rate of pruning by the SMT solver is not enough to 
offset the increased computational cost.
Thus, using the abstract interpreter strikes a good balance between
the precision and the computational cost.
As an example, for the benchmark {\tt hd-14-d1}, 
by the \textit{SMTSolver} variant, 
the SMT solver is invoked 2,247 times and used to prune 
1,091 partial programs. 
This takes 49.08 seconds out of 52.33 seconds spent for solving the benchmark.
On the other hand, 
by \tool,
the forward-backward analysis is invoked 1,230 times to prune 1,127 partial programs.
This takes only 0.06 second out of 0.15 second spent for solving the benchmark.
This shows the cost-effectiveness of using the abstract interpreter.

\paragraph{Summary of Results.}
Both of the forward and backward analyses are necessary to prune the search space effectively.
In addition, using the SMT solver is more expensive than using an abstract interpreter.






\section{Related Work}
\label{sec:related}

\paragraph{Synthesis with Abstract Interpretation.} 
Most of the previous pruning approaches to program synthesis by abstract interpretation have employed 
forward abstract interpretation only~\cite{simpl,scythe,morpheus,storyboard,synch,blaze}. 
Contrary to these approaches, 
our technique uses both forward and backward reasoning 
to derive necessary preconditions for missing expressions 
and use them to effectively prune the search space.

\citet{volt} accelerate synthesis via backward reasoning to check if 
necessary preconditions can be met by each candidate program.
In their setting, a necessary precondition for a partial program $P$ is a constraint on $P$'s
inputs that must be satisfied by any completion of $P$ in order
for $P$ to satisfy a given contraint over a desired new data structure. 
If the condition is falsified by any input, 
the partial program is discarded. 
Such a necessary precondition can be computed by the standard weakest precondition 
method. 
However, there is no synergistic combination of forward and backward reasoning, 
thereby potentially limiting the pruning power.

To the best of our knowledge, 
\citet{souper} is the only prior work that 
uses both forward and backward 
abstract interpretation to prune the search space of synthesis.  
However, a synergistic combination of forward and backward analyses 
is missing. 
In this work, 
the forward and backward analyses are performed separately  
without any interaction between them. 
In addition, their work is limited to LLVM superoptimization,
whereas our work is applicable to a wide range of inductive synthesis problems 
by targetting the SyGuS specification language. 

In contrast to our work that uses fixed abstract domains throughout the synthesis process, 
another line of work is based on abstraction refinement~\cite{blaze,blaze2,synch,tygar}. 
In these approaches, 
programs that do not satisfy the specification are used 
to iteratively refine the domain until a solution is found.
\blaze~\cite{blaze} with its extension~\cite{blaze2} automatically learns predicate abstract domains 
from a given set of predicate templates and training synthesis problems. 
This approach is useful when there is no expert having a 
good understanding of the target application domain. 
Our work shows that highly precise abstractions for abstract interpretation
can significantly improve the synthesis performance without abstraction refinement. 
We expect our key idea of using forward and backward analyses 
can be applied to abstraction refinement-based approaches
to further improve the synthesis performance.


\paragraph{Iterated Forward/Backward Analysis.}
The combination of forward and backward static analyses, which was first introduced in \cite{cousot:tel-00288657}, 
has been studied in the context of 
program verification~\cite{jlp92,10.1007/3-540-44978-7_7,10.1007/978-3-030-45234-6_9,Kanagasabapathi2020ForwardAB,10.1145/2678015.2682544}, 
model checking~\cite{cc99}, 
counterexample generation for failed specifications~\cite{10.1007/978-3-030-32304-2_13,10.1109/ICSE-Companion.2019.00067}, and  
filtering spurious static analysis alarms~\cite{10.1007/11547662_21}. 
In general, iterated forward and backward analyses increase the precision 
of the analysis at the cost of increased analysis time.

To the best of our knowledge, our work is the first to use the \emph{iterated} forward and backward analyses 
for inductive synthesis. 
Our key finding is that 
a highly precise analysis employing both forward and backward reasoning 
can increase the success rate of pruning, which is  
enough to offset the increased analysis time. 



\paragraph{Abstract Domains for Bit-Vector Arithmetic.} 

There is a large body of work on abstract domains for bit-vector arithmetic.
%
In the following, we briefly describe a few of them. 
\citet{mine:hal-00748094} and \citet{10.1145/1134650.1134657} 
proposed to use a combination of the interval domain and the bitwise domain.  
Some of our forward abstract transfer functions are borrowed from these works.
The wrapped interval domain~\cite{gange15} can precisely track effects of overflow and underflow 
by wrapping the bit-vector values around the minimum and maximum bit-vector values. 
\citet{10.1007/978-3-319-52234-0_27} proposed a framework for transforming 
numeric abstract domains over integers to bit-vector domains.
\citet{10.1007/978-3-540-74061-2_8} proposed a wrap-around operator for polyhedra 
to track the wrap-around effects.  

To the best of our knowledge, 
none of the existing domains for bit-vectors provide backward abstract transfer functions.
As already shown in our experiments, 
the backward abstract interpretation is crucial for pruning 
the search space. 
Therefore, to employ the existing domains for our approach,  
one needs to develop backward abstract transfer functions for them.

\paragraph{Domain Specializations for Inductive Synthesis.}
Recent works have demonstrated 
significant performance gains by exploiting 
domain knowledge in various forms 
such as domain-specific languages~\cite{flashfill,refazer,flashnormalize}, 
probabilistic models~\cite{probe,euphony}, 
inverse semantics~\cite{duet},
and templates~\cite{10.1007/978-3-319-40970-2_19}. 
In particular, 
\duet~\cite{duet} combines the bidirectional search with 
specialized \emph{inverse semantics} (also called witness functions) 
that return the set of possible inputs that can produce a given output. 
Our work combines the birectional search with abstract interpretation. 
Our experiments show that highly precise abstract semantics 
can provide significant performance gains 
that are complementary to those achieved by other domain specializations 
such as inverse semantics and probabilistic models, 
and it is promising to incorporate our approach into the previous approaches.

\section{Future Work}
\label{sec:futurework} 

A possible extension of our approach is to support 
other theories in SyGuS such as integer arithmetic and string theory. 
Due to insufficient precision of a single abstract domain, 
multiple abstract domains are necessary including forward/backward abstract transfer
functions and a reduction operator to define a reduced product of those abstract domains. 
For integer arithmetic, 
a reduced product of existing non-relational abstract domains such as 
the interval domain~\cite{cc77} and 
the congruence domain~\cite{gra89} can be used. 
Furthermore, 
relational domains~\cite{octagon,polyhedra} can aid in
tracking the relations between different program holes and input variables.  
For strings, 
abstract domains using
prefixes, suffixes, and simple regular expressions~\cite{costantini}, 
pushdown automata~\cite{stringpda},  
and parse stacks~\cite{doh} can be used.
Another possible extension is to synthesize programs with loops. 
In contrast to our current loop-free setting, 
the forward and backward analyses 
may require advanced widening/narrowing operators 
(e.g., widening with inferred thresholds~\cite{widening}) 
to expedite the convergence of the fixpoint iteration while maintaining precision.


\section{Conclusion}
\label{sec:conclusion}

We presented a novel program synthesis algorithm that effectively
prunes the search space by using a forward and backward abstract
interpretation. Our implementation \tool and its evaluation showed
that the performance is significantly better than the existing
state-of-the-art synthesis systems \duet and
\probe.  
The key to enable this
performance and scalability of inductive program
synthesis is to combine a forward abstract interpretation with a
backward one to rapidly narrow down the search space. Our experiments
also showed that using SMT solver on behalf of such sophisticated
static analysis does not scale.

\begin{acks}
  We thank the reviewers for insightful comments. 
  This work was supported by
    IITP (2022-0-00995)
  , NRF (2020R1C1C1014518, 2021R1A5A1021944)
  , Supreme Prosecutors' Office of the Republic of Korea
    grant funded by Ministry of Science and ICT(0536-20220043)
  , BK21 FOUR Intelligence Computing (Dept. of CSE, SNU) (4199990214639) 
    grant funded by the Korea government (MSIT)
  , Sparrow Co., Ltd.
  , Samsung Electronics Co., Ltd. (IO220411-09496-01)
  , Greenlabs (0536-20220078)
  , and Cryptolab (0536-20220081)
\end{acks}

\paragraph{Data-Availability Statement.} The artifact is available at Zenodo\cite{simba-artifact}.

\bibliographystyle{ACM-Reference-Format}
\bibliography{references}


\end{document}